\documentclass[12pt,a4paper]{article}

\usepackage{authblk}
\usepackage{amsmath,amssymb}
\usepackage{graphicx,psfrag,color}
\usepackage{amsfonts}
\usepackage{epsfig}
\usepackage{times,pstricks,pst-node}
\usepackage{pstricks,pst-node,pst-text,pst-3d}
\usepackage{caption}
\usepackage{setspace}
\usepackage[section]{placeins}

\renewcommand\footnotemark{}

\newcommand{\im}{\mathrm i}

\newcommand{\tr}{\operatorname{Tr}}
\newcommand{\str}{\operatorname{str}}

\newcommand{\ket}[1]{\left|#1\right\rangle}      % Ket-Zustand
\newcommand{\eq}{\begin{equation}}
\newcommand{\en}{\end{equation}}
\newcommand{\bear}{\begin{eqnarray}}
\newcommand{\ear}{\end{eqnarray}}

\begin{document}
\title{Exact thermodynamics and phase diagram of integrable t-J model with chiral interaction}
\author{T.S. Tavares$^*$\footnote{$^*$ tavares@df.ufscar.br}  and G.A.P. Ribeiro$^{\dagger}$ \footnote{$^{\dagger}$ pavan@df.ufscar.br}}

\affil{Departamento de F\'{i}sica, Universidade Federal de S\~ao Carlos \\ 13565-905 S\~ao Carlos-SP, Brazil}

\date{}
\maketitle
\setcounter{page}{1}
\thispagestyle{empty}

\begin{abstract}
We study the phase diagram and finite temperature properties of an integrable generalization of the one-dimensional super-symmetric t-J model containing interactions explicitly breaking parity-time reversal (PT) symmetries. To this purpose, we apply the quantum transfer matrix method which results in a finite set of non-linear integral equations. We obtain numerical solutions to these equations leading to results for thermodynamic quantities as function of temperature, magnetic field, particle density and staggering parameter. Studying the maxima lines of entropy at low but non zero temperature reveals the phase diagram of the model. There are ten different phases which we may classify in terms of the qualitative behaviour of auxiliary functions, closely related to the dressed energy functions.
\end{abstract}

\newpage

\section{Introduction}
Since the discovery of high-$T_c$ superconductors there have been many ideas to describe its pairing mechanism. The most prominent one is due to Anderson, who proposed that the insulating phase of superconducting copper oxides is described by a singlet spin liquid\cite{ANDERSON}. Further, the introduction of dopants would then result in pair of holes, which might form a condensate. An effective Hamiltonian to describe this problem is the two-dimensional Hubbard\cite{ANDERSON,ZHANG} or t-J model\cite{ZHANG}, depending on the relative magnitude of on-site Coulomb repulsion between holes at a Cu site and atomic energy levels of O holes\cite{ZHANG}. In the case of the former interaction  being  dominant, one has an appropriate description given by the t-J model. In both cases, however, the half-filling limit is described by the two-dimensional Heisenberg model on a square lattice.

This mechanism can lead to even more exotic type of superconductivity. In two spatial dimensions, fractional statistics may arise with macroscopic violation of parity ($P$) and time reversal ($T$) symmetries, while the system remains $PT$-symmetric\cite{WEN}. The spin liquid state related to the insulating phase would support a nonlocal extension of chiral order, hence a chiral spin liquid\cite{WEN,KALMEYER}. At first, this chiral spin liquid state was a candidate to the ground state of the two dimensional Heisenberg model on a square lattice with frustration introduced by next-nearest neighbor interaction\cite{WEN}. Nevertheless, this possibility was ruled out\cite{IMADA} and a rigorous proof of a spontaneous breaking $P$ and $T$ Hamiltonian exhibiting chiral spin liquid ground state is still missing\cite{ZVYAGINstg0,FRAHM2}. In absence of such spontaneous break of symmetry, one may include interactions explicitly breaking $P$ and $T$ to force the ground state into the elusive chiral spin
liquid. However, the above failure suggests that mean field and perturbative approaches are not reliable and the new state can only be tested on firm grounds, i.e, by means of exact solution.

Nevertheless, we lack exact results of two-dimensional (2D) quantum lattice models. Therefore, it should be interesting to investigate integrable quantum multi-chains interpolating between 1D and 2D behaviour\cite{ZVYAGINstg0}. This is possible thanks to the construction of staggered row-to-row transfer matrices with alternating spectral parameters. Notably, the resulting Hamiltonians are local with interactions explicitly breaking $PT$-symmetry. In \cite{ZVYAGINstg0} the 1D Heisenberg model was generalized to include next-to-nearest neighbour and chiral spin interactions. Further investigations were carried out to analyze its phase diagram and asymptotic behaviour of correlation functions\cite{FRAHM2}. Hence, the local chiral order in the ground-state spin liquid was exactly calculated\cite{MKHITARYAN}, although the chiral term is not a conserved quantity of the model.

Beyond the half-filling limit, there are also exact results concerning the t-J model. The condition for integrability and exact solution was provided in \cite{SCHLOTTMANN,KOREPIN} and asymptotic behaviour of correlation functions at zero temperature was calculated in \cite{KAWAKAMI}. This latter calculation confirmed Haldane's Luttinger liquid picture for metallic phases of 1D quantum systems\cite{HALDANE}. The same construction of \cite{ZVYAGINstg0} was employed to the t-J model, allowing for the exact solution of the multi-chain model and resulting in a conjecture of gapped excitations in absence of $PT$ breaking field\cite{ZVYAGINTJ}. Besides, in \cite{TJHIGHER} it was investigated the t-J model with a competing higher conserved quantity. It has been argued that commensurate to incommensurate phase transitions of charge and spin type takes place and also spontaneous magnetic and charge ordering.

In order to better understand the ground state phase diagram of the integrable t-J model with chiral interaction, it is required the exact computation of physical properties. To this end we use the quantum transfer matrix approach (QTM)\cite{MSUZUKI,KLUMPER92,DEVEGA0,KLUMPER93}, where free-energy of the 1D quantum model is mapped on the partition function of a 2D classical model. In this case, just the largest eigenvalue of the so called quantum transfer matrix contributes to the free-energy. The largest eigenvalue is given in terms of some auxiliary functions which are solutions of a finite set of non-linear integral equations(NLIE). This method revealed to be successful in describing thermodynamics of Heisenberg model and its generalization\cite{KLUMPER93,JSUZUKI,TRIPPE,RIBEIRO,TAVARES}, t-J model\cite{KLUMPER-TJ}, Hubbard model\cite{HUBBARD}, and $su(n|m)$ invariant models with $n+m < 5$ \cite{KLUMPER-SUN,DAMERAU}.

We address the above mentioned issue and apply the QTM approach to obtain a finite set of NLIE for the thermodynamics of multi-chain integrable generalization of t-J model containing chiral terms. Although the general kernel structure of these equations remain the same as compared with the usual t-J model \cite{KLUMPER-TJ}, the distinct physical behaviour is dictated by the new driving terms which appeared in the derivation of NLIE in this work. Our formalism also covers multi-chain models with arbitrary number of staggering parameters, however we focus on the study of the simplest case of two coupled chains. In this way, we are able to investigate its phase diagram and describe the quantum phases. We classify the phases exploiting the previous knowledge of the special limits $n\rightarrow 1$ (Heisenberg model with competing interactions) \cite{TAVARES}, $\theta=0$ (t-J model) \cite{KLUMPER-TJ} and also based in the qualitative behaviour of auxiliary functions, which is closely related to the dressed energy 
functions \cite{TJHIGHER}. In addition, we also study the case of high-magnetic fields, which allows us to obtain a good physical picture of the phase transitions of charge type in terms of free-fermions.

In order to describe the possible scenarios for the ground state diagram, we investigate physical quantities at low but finite temperature \cite{ZHU,TAVARES}. We emphasize that the phases we are studying are of quantum nature (quantum phase transitions) and occurs strictly at zero temperature, however we are able to describe the boundary phases at low but finite temperature. This is so, because at very low temperatures there exist a regime in which thermal fluctuations are insufficient to drive the system from its ground to excited states and therefore the quantum fluctuations dominate. So, the use of exact results at low but finite temperature contributes to the study of the possible scenarios of ground state phase diagram of quantum one-dimensional models.

This paper is organized as follows. First we review the construction\cite{ZVYAGINTJ} to obtain integrable models explicitly breaking  $PT$ symmetry and describe the QTM method to the thermodynamics\cite{KLUMPER93}. In Section \ref{non-linear}, we introduce suitable auxiliary functions and derive the NLIE describing the thermodynamics of the model. The section \ref{results} is devoted to the numerical solution of the NLIE leading to the phase diagrams of the two-coupled chain case. Finally, we summarize our results in  the section \ref{conclusion}. In the appendix we show the matching between the quantum phases of the model and qualitative behaviour of auxiliary functions.

\section{QTM approach to multi-chain generalizations of $su(n|m)$ invariant solution of Yang-Baxter equation}
\label{review}

The row-to-row transfer matrix has been identified as a generator of conserved currents\cite{BAX}. Therefore, we can obtain local Hamiltonians from the logarithmic derivative of the transfer matrix. The row-to-row transfer matrix of a staggered vertex model is given by
\eq
T(\lambda)=\str_{\cal A} \prod_{i=1}^{\stackrel{\curvearrowleft}{L}} \left[\prod_{k=1}^{\stackrel{\curvearrowleft}{M}} \mathcal{L}_{{\cal A}, (i-1) M+k}(\lambda,\im \omega_{M+1-k})\right] ,
\en
where the $\cal L$-operator $\mathcal{L}_{jk}(\lambda,\nu)$ acts non-trivially in two different $Z_2$ graded spaces, here assumed to have dimension $n+m$. Hence $\str_{\cal A}$ stands for super-trace on subspace ${\cal A}$ of graded matrices acting on ${\cal A}\otimes V_1\otimes\cdots\otimes  V_{M L}$\cite{CORNWELL}. The $M$ staggering parameters $\omega_k$ introduce inhomogeneity along the row in a way that the model remains invariant by cyclic translation of $M$ sites.
Each local space $V_j$ (${{\cal A}}$) supports a Grassmann field, nevertheless we require $\cal L$-operators to be complex and homogeneous even element of $End(V \otimes V)$.
In our notation, the $\cal L$-operators are given by
\eq
\mathcal{L}(\lambda,\nu)=\sum_{\alpha,\beta,\gamma,\delta} \check{\mathcal{L}}_{\alpha,\gamma}^{\beta ,\delta}(\lambda,\nu) e_{\alpha\beta}^{(j)} e_{\gamma\delta}^{(k)},
\en
where $e_{\alpha\beta}^{(j)} =\mbox{Id} \stackrel{s}{\otimes} \ldots \underbrace{e_{\alpha\beta}}_j\ldots \stackrel{s}{\otimes} \mbox{Id}$, $e_{\alpha\beta}$ are the Weyl matrices and $\stackrel{s}{\otimes}$ denotes the super-tensor product\cite{SKLYANIN}. This operator satisfies the graded Yang-Baxter equation (GYBE)\cite{SKLYANIN}
\eq
{\cal L}_{12}(\lambda, \mu){\cal L}_{13}(\lambda, \gamma){\cal L}_{23}(\mu, \gamma)={\cal L}_{23}(\mu, \gamma){\cal L}_{13}(\lambda, \gamma){\cal L}_{12}(\lambda, \mu).
\en
We also restrict ourselves to solutions of GYBE that have the following symmetry properties:
\begin{align}
  \mbox{Regularity: } & {\cal L}_{12}(\lambda, \lambda)=P_{12}^g, \label{regul} \\
  \mbox{Unitarity: } & {\cal L}_{12}(\lambda,\mu) {\cal L}_{12}(\mu,\lambda)= \mbox{Id}, \label{uni} \\
  \mbox{Time reversal: } & {\cal L}_{12}^{st_1}(\lambda, \mu)={\cal L}_{12}^{st_2}(\lambda, \mu), \label{time-rev}
\end{align}
where $P_{12}^g$ is the graded permutation operator and $st_k$ denotes the super-transposition on the $k$-th space\cite{CORNWELL,SKLYANIN}. Thanks to GYBE we have the commutativity property of the row-to-row transfer matrices
\eq
[T(\lambda),T(\mu)]=0,\qquad \forall \lambda,\mu.
\label{comm}
\en
Under these circumstances, $T(\lambda)$ has many conserved quantities. The properties of regularity and unitarity ensures that one may obtain local operators from the logarithmic derivative of the transfer matrix. To this end, we consider $M$ trivially related transfer matrices $T_j(\lambda;\im\vec{\omega})=T(\lambda+\im \omega_j;\im \vec{\omega})$, such that\cite{TAVARES}
\eq
t(\lambda)=\prod_{j=1}^M T_j(\lambda;\im\vec{\omega}).
\label{transfer-prod}
\en
Therefore one finds
\bear
{\cal H(\vec{\omega})}&=&\frac{1}{M}\frac{d}{d \lambda}\ln{t(\lambda)}\Big|_{\lambda=0}, \\
	&=&\frac{1}{M} \sum_{q=0}^{M-1}{\rm e}^{- \im q\mathcal{P}} {\cal H}_1(\vec{\omega}_{+q}) {\rm e}^{ \im q{\cal P}},
 \label{Hgen}
\ear
where ${\cal P}$ is the momentum governing one-site cyclic translation to the right,
\begin{multline}
 {\cal H}_1(\vec{\omega})=\sum_{i=1}^L \sum_{k=1}^M \left[\prod_{n=1}^{M-k}{\cal L}_{Mi+n,Mi}(\im\omega_{M+1-n},\im \omega_1)\right]{\cal L}_{M(i+1)-k+1,Mi}(\im \omega_{k},\im \omega_1) \times \\ \times \frac{d}{d \lambda} {\cal L}_{Mi,M(i+1)-k+1}(\lambda+\im \omega_1,\im \omega_k)\bigg|_{\lambda=0}\left[\prod_{n=1}^{\stackrel{\curvearrowleft}{M-k}}{\cal L}_{Mi,Mi+n}(\im \omega_1,\im \omega_{M+1-n})\right], \label{H1}
 \end{multline}
and we have introduced the notation\cite{TAVARES}
\eq
\vec{\omega}_{+q}=(\omega_{1+q},\omega_{2+q},\ldots, \omega_{M+q}), \qquad \vec{\omega}\equiv \vec{\omega}_{+0},~~\omega_{k+M}\equiv\omega_{k}.
\nonumber
\en
The simplest solution of GYBE is the so called Perk-Schultz model\cite{PERK}, where
\eq
{\cal L}(\lambda,\mu)= {\cal L}(\lambda -\mu,0)=\frac{(\lambda-\mu)\mbox{Id}+P^g}{1+\lambda-\mu}.
\en
The difference property of this ${\cal L}$-operator allow us to fix $\omega_k=\theta_{k-1}$ and $\theta_0=0$, without any loss of generality. Here we shall be interested in the solution with $(n,m)=(2,1)$.

Apart from trivial additive factors, the solution with $M=1$ gives rise to the known super-symmetric t-J model\cite{SCHLOTTMANN}:
\eq
\mathcal{H}^{\text{t-J}}=\sum_{j=1}^L h^{\text{t-J}}_{jj+1}=\sum_{j=1}^L  \left(-\sum_{\tau}( c^\dag_{j+1 \tau} c_{j \tau}+c^\dag_{j \tau} c_{j+1 \tau})+ 2 \vec{S}_j \cdot \vec{S}_{j+1}-\frac{1}{2}\sum_{\sigma \tau} n_{j \tau} n_{j+1 \sigma}\right),
\label{tjsimp}
\en
 where $n_{j \tau}=c_{j \tau}^{\dag} c_{j \tau}$, $S_j^k=\sum_{\tau\sigma} S^k_{\tau \sigma} c_{j \tau}^{\dag} c_{j \sigma}$ ($k=x,~y,~z$) and
$c_{j \tau}$ are the ``projected'' fermionic operators acting on subspace $\ket{\uparrow}, ~\ket{0}, ~\ket{\downarrow}$ with grading $\{0, ~1, ~0\}$. These operators satisfy the following anti-commutation rules\cite{CHAO}
\eq
\left[c_{i \tau},c_{j \sigma}\right]_+=[c_{i \tau}^\dag,c_{j \sigma}^\dag]_+=0, \nonumber
\en
\eq
[c_{i \tau},c_{j \sigma}^\dag]_+=((1-n_{i-\tau})\delta_{\tau \sigma}+S_i^{-\tau}(1-\delta_{\tau \sigma}))\delta_{ij}.
\en
The simplest multi-chain generalization (\ref{Hgen}, \ref{H1}) of the super-symmetric t-J model occurs for $M=2$ and was obtained in \cite{ZVYAGINTJ} with chiral interactions written in terms of Weyl matrices. We write this Hamiltonian explicitly in terms of fermionic operators $c_{i \tau}$
{\small
\begin{multline}
\mathcal{H}^{\text{t-J}}(\theta)=\frac{1}{2(1+\theta^2)}\sum_{j=1}^{2 L} 2 h^{\text{t-J}}_{jj+1}+\theta^2 h^{\text{t-J}}_{jj+2}+ {(-1)}^j 4 \theta \vec{S}_j \cdot \vec{S}_{j+1} \times \vec{S}_{j+2}+\\
{(-1)}^j 4 \theta \sum_{p\{j,j+1,j+2\}}{(-1)}^{\mbox{sgn}(p)}\left(\vec{s}_{p(j) p(j+1)} \cdot \vec{S}_{p(j+2)}+m_{p(j) p(j+1)}(1-\sum_\tau\frac{n_{p(j+2) \tau}}{2})\right),
\label{Htjgen}
\end{multline}}where the sum over ${p\{j,j+1,j+2\}}$ denotes summation over cyclic permutation of indices $j,~j+1,~j+2$ with $\mbox{sgn}(p)$ the usual signature of permutations. Hence, we have the ``de-localized'' analogues of particle density and spin operators:
\eq
m_{jk}=\frac{\im}{4} \sum_\tau c_{k \tau}^\dag c_{j \tau}-c_{j \tau}^{\dag}c_{k \tau},
\en
\eq
\vec{s}_{jk}=\{{s}_{jk}^x,{s}_{jk}^y,{s}_{jk}^z\}=\left\{\frac{s_{jk}^++s_{jk}^-}{2},\frac{s_{jk}^+-s_{jk}^-}{2\im},{s}_{jk}^z\right\}, \nonumber
\en
\bear
&2{s}_{jk}^+=\im(c_{k\uparrow}^\dag c_{j\downarrow}-c_{j\uparrow}^\dag c_{k\downarrow}), \qquad 2{s}_{jk}^-=\im(c_{k\downarrow}^\dag c_{j\uparrow}-c_{j\downarrow}^\dag c_{k\uparrow}), \nonumber \\
&4 {s}_{jk}^z=\im(c_{k\uparrow}^\dag c_{j\uparrow}-c_{j\uparrow}^\dag c_{k\uparrow}-c_{k\downarrow}^\dag c_{j\downarrow}+c_{j\downarrow}^\dag c_{k\downarrow}).
\ear
We have neglected additive constants contributing to the zero of energy and terms proportional to the particle density. In what follows those terms can be controlled by introducing generalized chemical potentials in the calculation of partition function.

We are interested in the partition function of the quantum model $Z=\tr[e^{-\beta {\cal H}}]$, which can be obtained by the Trotter-Suzuki decomposition. This is done by noting that the transfer matrix can be written as
\eq
t(\lambda)=e^{M \im {\cal P}+\lambda M \mathcal{H} +O(\lambda^2)},
\label{texp}
\en
and the conjugated transfer matrix is given by
\eq
\bar{t}(\lambda)=\prod_{j=1}^M \bar{T}_j(\lambda,\im\vec{\omega}),
\label{bartransfer}
\en
where $\bar{T}_j(\lambda,\im\vec{\omega})=\bar{T}(\lambda-\im \omega_j,\im \vec{\omega})$ and
\eq
\bar{T}(\lambda,\im \vec{\omega})=\str_{\cal A}{\left[\prod_{i=1}^{\stackrel{\curvearrowleft}{L}} \left[\prod_{k=1}^{\stackrel{\curvearrowleft}{M}} \mathcal{L}_{{\cal A},(i-1) M+k}^{st_{\cal A}}(\im \omega_{M+1-k},-\lambda)\right]  \right]}.
\label{transferBAR}
\en
Therefore, the conjugated transfer matrix $\bar{t}(\lambda)$ can also be written as 
\eq
\bar{t}(\lambda)=e^{-M \im {\cal P}+\lambda M \mathcal{H} +O(\lambda^2)}.
\label{tbarexp}
\en
The product $t(\lambda)\bar{t}(\lambda)$ contain the Hamiltonian as the leading term in the exponent. By choosing $\lambda=-\frac{\beta}{M N}$ we have that terms in $O(\lambda^2)$ becomes small compared to $O(\lambda)$ as the Trotter number $N$ goes to infinity. Thus we have
\eq
Z=\lim_{N\rightarrow\infty}\tr{\left[(t(-\tau)\bar{t}(-\tau))^{N/2} {\rm e}^{\beta \sum_{j=1}^{n+m}\mu_j \hat{N}_j}\right]}=\tr{{\rm e}^{-\beta (\cal{H}-\mu \hat{N})}},  \qquad \tau=\frac{\beta}{M N},
\label{particao}
\en
where for $su(n|m)$ invariant models the number of each particle species $\hat{N}_j=\sum_{k=1}^{M L} n_{j k}$, with $j=1,\ldots,n+m$, is a conserved quantity. The operator $n_{j k}$ gives $1$ if it acts on a state with a particle of species $j$ at site $k$ and $0$ otherwise. Hence, depending on application, one is allowed to evaluate the canonical or grand-canonical partition function of the quantum chain as long as one can evaluate the partition function of a staggered vertex model with the suitable twisted boundary condition. Here we set $\mu_{1}=\frac{H}{2}+\mu$, $\mu_{2}=-\sum_{j=1}^{M}{\frac{1}{1+\theta_{j-1}^2}}$, $\mu_{3}=-\frac{H}{2}+\mu$ accounting for the grand-partition function of model (\ref{Htjgen}) in a external magnetic field.

The column-to-column transfer direction is more appropriate to evaluate the partition function (\ref{particao}). Therefore we define the quantum transfer matrix
\begin{multline}
t^{QTM}(x)=\tr_{Q}\Bigg[ \prod_{i=1}^{\frac{N}{2}} ~\left[\prod_{j=1}^M {\cal L}_{ M(2i-2)+j,Q}(-\tau+ \im \omega_j, -\im x)\right]\times\\ \times \left[\prod_{j=1}^M{\cal L}_{ M (2i-1)+j,Q}^{st_Q}(-\im x,\tau+ \im \omega_j)\right]\Bigg],
\label{qtm-gen}
\end{multline}
where the new spectral parameter $x$ was introduced to guarantee the commutativity property of the quantum transfer matrix $[t^{QTM}(x),t^{QTM}(x')]=0$ and renders the QTM integrable.

The (grand-)partition function can be written in terms of the quantum transfer matrix (\ref{qtm-gen}) as follows
\eq
Z=\lim_{N\rightarrow\infty}\str{\left[\prod_{j=1}^M (t^{QTM}(- \omega_j))^{L}\right]}.
\en
This allow us to express the thermodynamic potential in terms of the largest eigenvalue $\Lambda_{max}^{QTM}(x)$ of the quantum transfer matrix,
\bear
f&=&-\frac{1}{\beta}\lim_{L,N\rightarrow \infty}\frac{1}{M L} \ln{Z},\\
 &=&-\frac{1}{\beta}\lim_{N\rightarrow \infty} \frac{1}{M}\sum_{j=1}^M\ln{\Lambda_{max}^{QTM}(- \omega_j)}.
 \label{free-energy}
\ear

\section{NLIE for the multi-chain generalization of super-symmetric t-J model}\label{non-linear}

The computation of the physical properties (\ref{free-energy}) requires the knowledge of the largest eigenvalue of the quantum transfer matrix in the infinity Trotter number limit. This is efficiently obtained by the quantum transfer matrix approach.

The quantum transfer matrix (\ref{qtm-gen}) can be diagonalized by Bethe ansatz techniques\cite{KULISH}. For instance, the eigenvalues can be obtained by the algebraic Bethe ansatz\cite{EKS,GOHMANN,RIBEIRO06}. This way, we have the following expression for the QTM eigenvalues of multi-chain generalization of $su(n|m)$ invariant models
\eq
\Lambda^{QTM}(x)=\sum_{j=1}^{n+m} \lambda_j(x), \qquad \lambda_j(x)={\rm e}^{\beta \mu_j} X(x) \prod_{k=1}^{m_{j-1}} \frac{a_{\epsilon_j}(\im x-\im x_k^{j-1})}{b(\im x-\im x_k^{j-1})} \prod_{k=1}^{m_{j}} \frac{a_{\epsilon_j}(\im x_k^{j}-\im x)}{b(\im x_k^{j}-\im x)},
\en
\eq
X(x)={\left[\prod_{k=1}^M b(-\tau+\im (x+\theta_{k-1})) b(-\tau-\im (x+\theta_{k-1}))\right]}^{\frac{N}{2}},
\en
where $a_{\epsilon_j}(x)=\frac{x+\epsilon_j}{x+1}$, $b(x)=\frac{x}{1+x}$ and $\epsilon_j={(-1)}^{p_j}$ with $p_j=0,~1$ the grading choices of state $j$. Also $x_k^0=-\im \tau -\theta_{\mbox{mod}(k-1,M)}$ and $x_k^{n+m}=\im \tau -\theta_{\mbox{mod}(k-1,M)}$ with $m_0=m_{n+m}=\frac{N M}{2}$. The Bethe ansatz equations reads
\eq
 \frac{{\rm e}^{\beta \mu_j}\prod_{k=1}^{m_{j-1}}\frac{a_{\epsilon_j}(\im x_r^j-\im x_k^{j-1})}{b(\im x_r^j-\im x_k^{j-1})}}{{\rm e}^{\beta \mu_{j+1}}\prod_{k=1}^{m_{j+1}}\frac{a_{\epsilon_{j+1}}(\im x_k^{j+1}-\im x_r^{j})}{b(\im x_k^{j+1}-\im x_r^{j})}}=\epsilon_j \epsilon_{j+1} \prod_{\stackrel{k=1}{k \neq r}}^{m_j} \frac{b(\im x_k^j-\im x_r^j) a_{\epsilon_{j+1}}(\im x_r^j-\im x_k^j)}{a_{\epsilon_j}(\im x_k^j-\im x_r^j)b(\im x_r^j-\im x_k^j)},
\en
with $j=1,\ldots,n+m-1$ and $r=1,\ldots,m_j$.

In order to take the Trotter limit, one has to define suitable auxiliary functions, exploit its analyticity properties and encode this Bethe ansatz roots information in a system of integral equations. For $su(2|1)$ case, the auxiliary functions are given in terms of the building blocks $\lambda_1(x),~\lambda_2(x),~\lambda_3(x)$ along the same lines as in \cite{KLUMPER-TJ}. Nevertheless, we additionally perform a particle-hole transformation on the $\mathfrak{c}(x)$ function, which results in a simpler half-filling limit. Our auxiliary functions are given as
\begin{align}
\mathfrak{b}(x)&=\frac{\lambda_1(x+\frac{\im}{2})}{\lambda_2(x+\frac{\im}{2})+\lambda_3(x+\frac{\im}{2})}=\frac{{\rm e}^{\beta \mu_1}}{{\rm e}^{\beta \mu_2}+{\rm e}^{\beta \mu_3}}\frac{\Phi_+(x-\frac{\im}{2}) \Phi_-(x+\frac{\im}{2}) q_1(x+\frac{3 \im}{2})}{\Phi_+(x+\frac{\im}{2}) q_2^h(x+\frac{\im}{2}) q_2(x-\frac{\im}{2})}, \nonumber
\\
\bar{\mathfrak{b}}(x)&=\frac{\lambda_3(x-\frac{\im}{2})}{\lambda_1(x-\frac{\im}{2})+\lambda_2(x-\frac{\im}{2})}=\frac{{\rm e}^{\beta \mu_3}}{{\rm e}^{\beta \mu_1}+{\rm e}^{\beta \mu_2}}\frac{\Phi_+(x-\frac{\im}{2}) \Phi_-(x+\frac{\im}{2}) q_2(x-\frac{3 \im}{2})}{\Phi_-(x-\frac{\im}{2}) q_1^h(x-\frac{\im}{2}) q_1(x+\frac{\im}{2})}, \nonumber
\\
\mathfrak{c}(x)&=\frac{\lambda_2(x) (\lambda_1(x)+\lambda_2(x)+\lambda_3(x))}{\lambda_1(x) \lambda_3(x)}={\rm e}^{\beta (\mu_2-\mu_1-\mu_3)}\Lambda(x),
\label{auxlower}
\end{align}
where $q_0(x)=\Phi_{+}(x)={\left[\prod_{j=1}^M\left( x-\theta_{j-1}+\im \tau\right)\right]}^{\frac{N}{2}}$, $q_{1,2}(x)=\prod_{k=1}^{m_{1,2}}(x-x_j^{1,2})$, $q_3(x)=\Phi_{-}(x)={\left[\prod_{j=1}^M\left( x-\theta_{j-1}-\im \tau\right)\right]}^{\frac{N}{2}}$   and $q_{1,2}^h(x)=\prod_{k=1}^{m_{1,2}}(x-{x_j^{h 1,2}})$ contains the hole solutions of Bethe ansatz equations, providing factorization property in the form
\eq
\frac{\lambda_j(x)+\lambda_{j+1}(x)}{{\rm e}^{\beta \mu_j}+{\rm e}^{\beta \mu_{j+1}} }=X(x)\frac{q_j^h(x)(\delta_{\epsilon_j ,\epsilon_{j+1}}+\delta_{\epsilon_j ,-\epsilon_{j+1}} . q_j(x+\im \epsilon_j))}{q_{j-1}(x)q_{j+1}(x)}.
\en
We locate the QTM largest eigenvalue in sector $m_1=m_2=\frac{M N}{2}$. For $H=0$, the Bethe Ansatz roots $x_k^{1}$($x_k^{2}$) have imaginary part distributed along a slightly deformed line above(below) from $\Im(z)=0$ without crossing lines  $\Im(z)=\frac{1}{2}$($\Im(z)=-\frac{1}{2}$). Similarly, the hole solutions $x_k^{h1}$($x_k^{h2}$) are distributed along lines slightly deformed from $\Im(z)=1(-1)$ without crossing lines  $\Im(z)=\frac{3}{2}$($\Im(z)=-\frac{3}{2}$). In this sense, the role of $\theta_j$ parameters is to produce deformations of Bethe roots along these lines and do not change the analyticity strip. By introducing magnetic field, there occurs some  vertical displacement in root patterns, although insufficient to violate analyticity hypothesis. Therefore, auxiliary functions (\ref{auxlower}) are analytical and non zero in a strip containing the real axis, with constant asymptotics. We also should mention that $\Lambda(x)$, for the largest eigenvalue, has an analytical
non zero strip at least containing $-\frac{1}{2}\leq \Im(z)\leq \frac{1}{2}$.

Moreover, we introduce some additional functions given by
\begin{align}
\mathfrak{B}(x)&=1+\mathfrak{b}(x)=\frac{\Lambda(x+\frac{\im}{2})}{{\rm e}^{\beta \mu_2}+{\rm e}^{\beta \mu_3}} \frac{\Phi_+(x-\frac{\im}{2}) \Phi_-(x+ \frac{3 \im}{2}) q_1(x+\frac{\im}{2})}{\Phi_+(x+\frac{\im}{2}) q_2^h(x+\frac{\im}{2}) q_2(x-\frac{\im}{2})}, \nonumber
\\
\bar{\mathfrak{B}}(x)&=1+\mathfrak{b}(x)=\frac{\Lambda(x-\frac{\im}{2})}{{\rm e}^{\beta \mu_1}+{\rm e}^{\beta \mu_2}} \frac{\Phi_+(x- \frac{3\im}{2}) \Phi_-(x+ \frac{\im}{2}) q_2(x-\frac{\im}{2})}{\Phi_-(x-\frac{\im}{2}) q_1^h(x-\frac{\im}{2}) q_1(x+\frac{\im}{2})}, \nonumber
\\
\mathfrak{C}(x)&=1+\mathfrak{c}(x)=({\rm e}^{\beta \mu_1}+{\rm e}^{\beta \mu_2}) ({\rm e}^{\beta \mu_2}+{\rm e}^{\beta \mu_3}) \frac{{\rm e}^{-\beta (\mu_1+\mu_3)} q_1^h(x) q_2^h(x)}{\Phi_+(x-\im) \Phi_-(x+\im)},
\end{align}
which also have an analytical non zero strip containing the real axis. The simply related functions $\mathfrak{B}(x)$ and $\mathfrak{b}(x)$ (likewise to the other functions) contains the same information, although they differ in their asymptotic behaviour.

The analyticity properties of the above auxiliary functions allow us to apply the Fourier transform on their logarithmic derivative. This results in a set of algebraic equation in the Fourier space for the auxiliary functions. The solution of this algebraic equations allows the limit $N \rightarrow \infty$ to be performed analytically and the final results can be transformed back to real space, resulting
\begin{align}
\ln \mathfrak{b}(x)&=-\beta K_{\theta}(x)+\beta \frac{H}{2}+F \ast \ln \mathfrak{B}(x)-F \ast \ln \mathfrak{\bar{B}}(x+\im)-K \ast \ln \mathfrak{C}(x), \nonumber\\
\ln \bar{\mathfrak{b}}(x)&= -\beta K_{\theta}(x)-\beta \frac{H}{2}-F \ast \ln \mathfrak{B}(x-\im)+F \ast \ln \mathfrak{\bar{B}}(x)-K \ast \ln \mathfrak{C}(x),\nonumber\\
\ln \mathfrak{c}(x)&= \beta F_{\theta}(x)-\beta \mu'+K\ast\left(\ln \mathfrak{B}(x)+\ln \bar{\mathfrak{B}}(x)\right)+F\ast \ln \mathfrak{C}(x),
\label{nlieq}
\end{align}
where $\mu'=\mu-\mu_2$, $K(x)=\frac{\pi}{\cosh(\pi x)}$, $F(x)=\int_{-\infty}^{\infty}\frac{{\rm e}^{\im k x}}{1+{\rm e}^{|k|}} {\rm d}k$ and $R_{\theta}(x)=\frac{1}{M}\sum_{j=1}^M R(x+\theta_{j-1})$, where $R(x)$ stands for an arbitrary function. The symbol $*$ denotes the convolution $f*g(x)=\frac{1}{2 \pi}\int_{-\infty}^{\infty} f(x-y)g(y)dy$. Equations (\ref{nlieq}) are a self-consistent set of non linear integral equations which describes the many-body integrable Hamiltonian (\ref{Htjgen}). Once we solve these, we may obtain the eigenvalue $\Lambda(x)$ or, by virtue of (\ref{free-energy}), the thermodynamic potential
\eq
f=-\mu+e_0-\frac{1}{M \beta}\sum_{j=1}^M K \ast \left(\ln \mathfrak{B}(-\theta_{j-1})+\ln \bar{\mathfrak{B}}(-\theta_{j-1})\right)-F\ast \ln\mathfrak{C}(-\theta_{j-1}),
\en
where
\eq
e_0=-\int_{-\infty}^{\infty} \left[ \frac{1}{1+{\rm e}^{|k|}}\right] {\Bigg|\frac{\sum_{j=1}^M {\rm e}^{\im k \theta_{j-1}}}{M} \Bigg|}^2{\rm d}k,
\nonumber
\en
is the ground-state energy at half-filling and zero external magnetic field.

\section{Numerical results}\label{results}

We can solve the NLIE by iteration where the convolutions are calculated in Fourier space by Fast Fourier Transform algorithm. In this way, we compute the thermodynamic potential as a function of temperature, magnetic field and chemical potential.

Another key procedure to avoid numerical differentiations in the computation of physical quantities e.g entropy and specific heat is to derive additional integral equations by differentiating the NLIE with respect to temperature, magnetic field and chemical potential. This implies in the following relations among the capital and lower case auxiliary functions\cite{KLUMPER-TJ}
\bear
\partial_r \log A=\frac{a}{A} \partial_r \log a, \qquad \partial_{r~s}^2 \log A=\frac{a}{A} \left(\partial_{r~s}^2 \log a-\frac{\partial_r \log a \partial_s \log a}{A}\right),
\ear
where $r$($s$) is any of $T,~\mu$ or $H$ and $a$($A$) is any lower-case(capital) auxiliary functions.

With the above expressions all derivatives up to second order of thermodynamical potential may be obtained as a function of $T,~\mu$ and $H$. However, we would like to eliminate $\mu$ in favor of $n$, the particle density. This is done by means of Newton method. Since compressibility ${\kappa}_H=\Big(\frac{\partial n}{\partial \mu}\Big)_{H,T}$ is at our disposal, we can set the particle density and find the corresponding $\mu$ within some required precision. Therefore, we can study thermodynamical properties as function of $n,~T$ and $H$.

Our interest is to describe the $n-\theta-H$ 3D diagram of model (\ref{Htjgen}). We analyze the entropy at very low but non-zero temperature. By virtue of thermodynamics fundamental laws we have vanishing entropy at $T=0$. Nevertheless, at low but finite temperatures the entropy is non vanishing and accumulates close to quantum phase transitions\cite{ZHU}. Although the absolute value of entropy is small, one may use it to trace the phase diagram since the maxima peaks becomes sharper as we approach the phase transition. Therefore, we can describe, at low but finite temperature, the possible scenarios for the ground state phase diagram, whose different phases are listed in Table 1.

\begin{table}[t!]
\begin{center}
\begin{tabular}{|c|c|c|}
  \hline
   Phase &  Spin  & Charge   \\ \hline
 \textrm{I} & commensurate anti-ferro   & commensurate metal   \\ \hline
 \textrm{II} & commensurate anti-ferro   &incommensurate metal    \\ \hline
 \textrm{III} & commensurate anti-ferro   & insulating \\ \hline
 \textrm{IV} & incommensurate anti-ferro   & commensurate metal    \\ \hline
 \textrm{V} & incommensurate anti-ferro   & incommensurate metal    \\ \hline
 \textrm{VI} & incommensurate anti-ferro   & insulating  \\ \hline
 \textrm{VII} & ferro & commensurate metal    \\ \hline
 \textrm{VIII}& ferro & incommensurate metal   \\  \hline
 \textrm{IX} & ferro & insulating  \\ \hline
 \textrm{X} & zero density & zero density \\ \hline
\end{tabular}
\caption{Phase Classification}
\label{table1}
\end{center}
\end{table}

We start describing two special limits, the $H-\theta$ plane at $n=1$ which is the Heisenberg model with competing interactions\cite{TAVARES} and the $n-H$ plane at $\theta=0$ which is the usual t-J model\cite{KLUMPER-TJ}. 

\begin{figure}[t!]
\begin{minipage}{0.5\linewidth}
\begin{center}
\includegraphics[width=\linewidth]{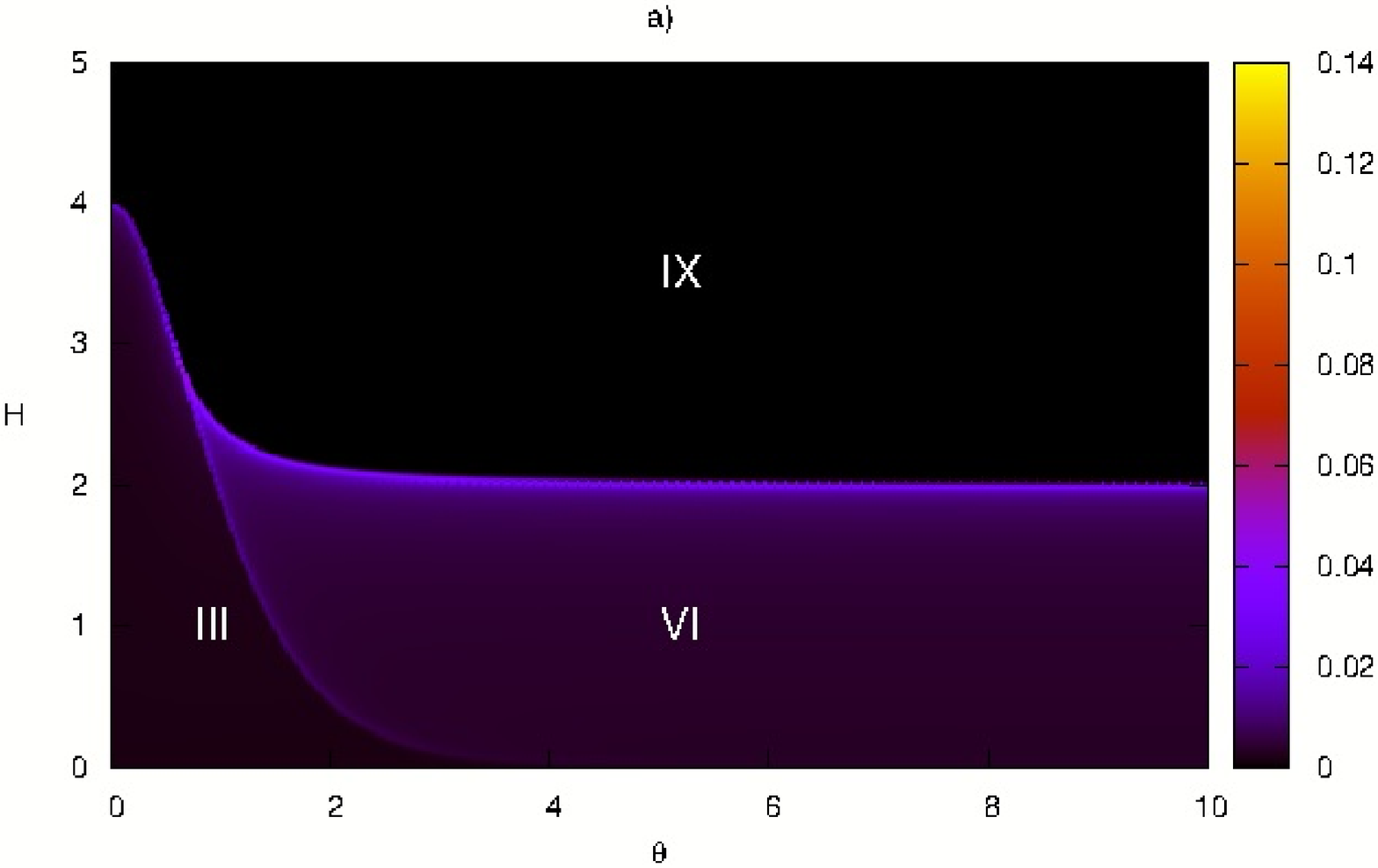} 
\end{center}
\end{minipage}%
\begin{minipage}{0.5\linewidth}
\begin{center}
\includegraphics[width=\linewidth]{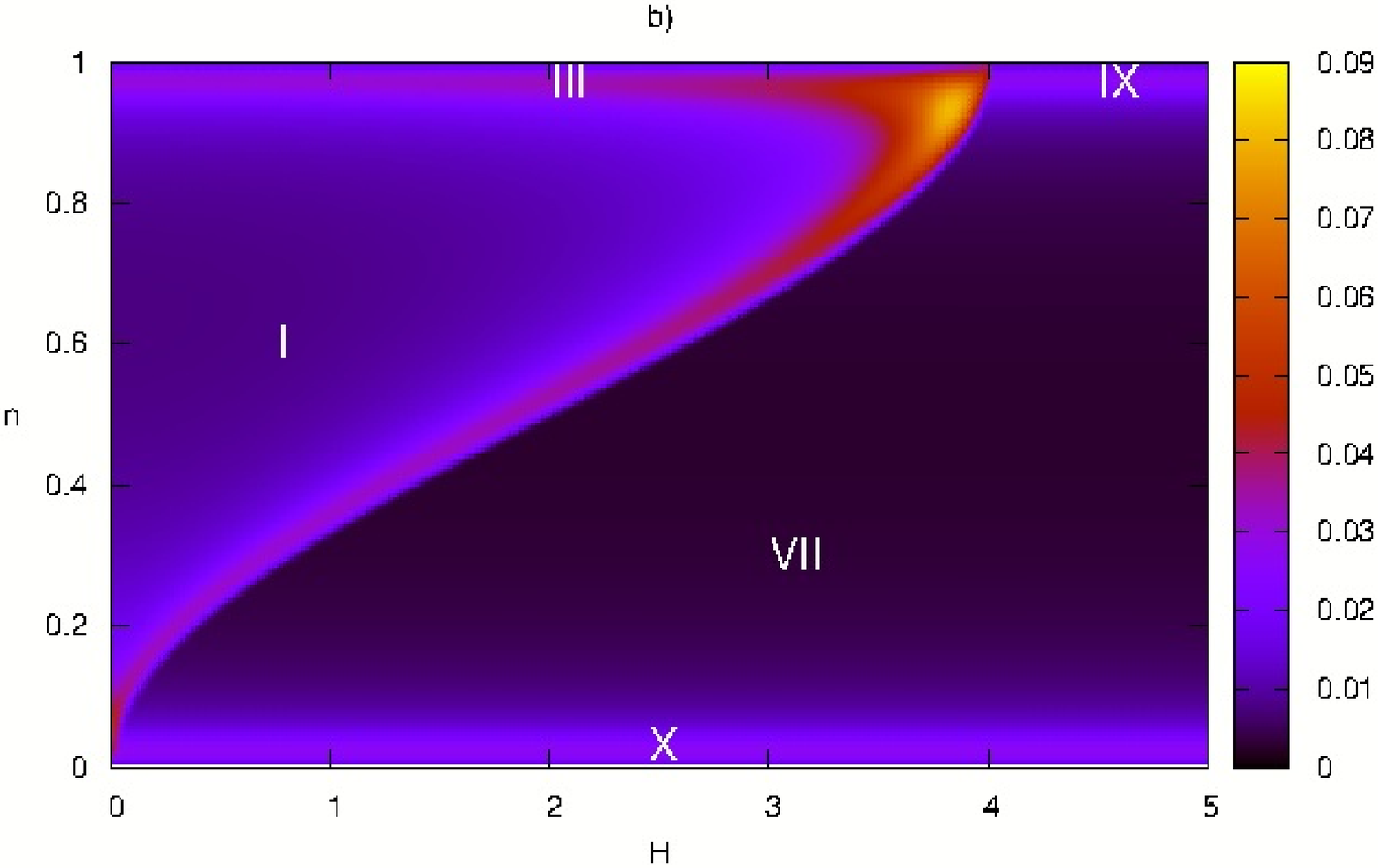} 
\end{center}
\end{minipage}
\caption{(Color online) Phase diagram obtained from entropy profile at $T=0.005$: a) half-filling ($n \rightarrow 1$); b) single-chain ($\theta \rightarrow 0$) limit.}
\label{fig01}
\end{figure}

The first case is shown in Figure \ref{fig01}a and refers to the limit $\mu \rightarrow \infty$. This case was first considered in \cite{ZVYAGINstg0,FRAHM2} and consists of an insulating phase with three different magnetic orders ranging from anti-ferromagnetic commensurate \textrm{III}, incommensurate \textrm{VI} to ferromagnetic \textrm{IX} order. The solution of the non-linear integral equations for the thermodynamics was proved to be a useful tool to determine this phase diagram\cite{TAVARES}. Moreover, the auxiliary functions analysis are closely related to dressed-energy functions analysis in $T \rightarrow 0$ limit\cite{KLUMPER-TJ}, which provide us with additional information about the nature of the phase transitions\cite{TRIPPE}, see Appendix. 

It is worth to note that the above non-linear integral equations (\ref{nlieq}) defined by means of a particle-hole transformation in the function $\mathfrak{c}$ have the advantage in relation to the equation for the usual t-J model introduced in \cite{KLUMPER-TJ} that the limit $n \rightarrow 1$ is obtained naturally. This occurs since the modified $\mathfrak{c}$ function behaves as ${\rm e}^{-\beta \mu'}$ in this limit and, therefore, convolutions containing $\ln \mathfrak{C}$ vanish and leave us with the equations of the Heisenberg model with competing interactions\cite{TAVARES}.

\begin{figure}[t!]
\begin{minipage}{0.32\linewidth}
\begin{center}
\includegraphics[width=\linewidth]{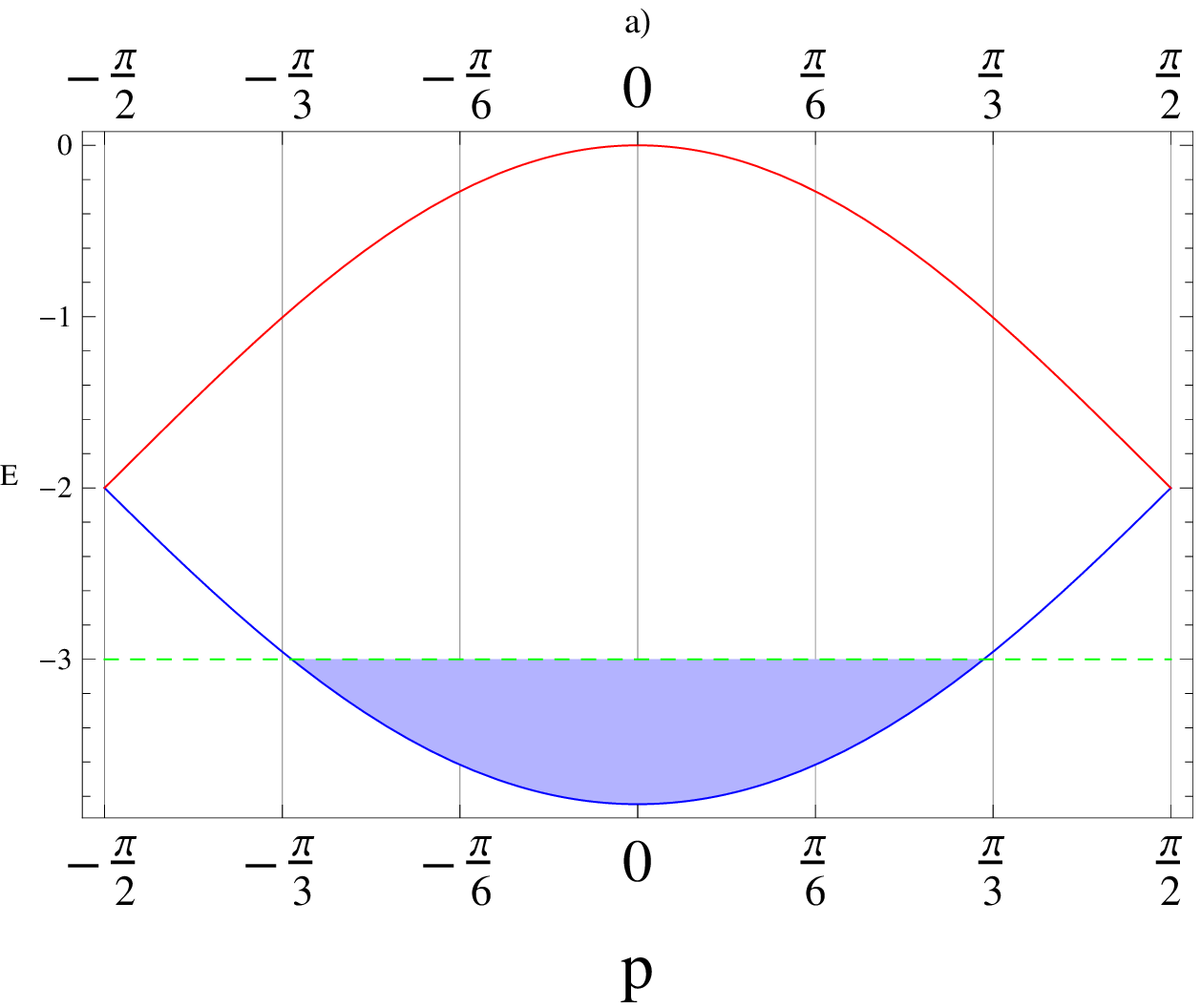} 
\end{center}
\end{minipage}%
\begin{minipage}{0.32\linewidth}
\begin{center}
\includegraphics[width=\linewidth]{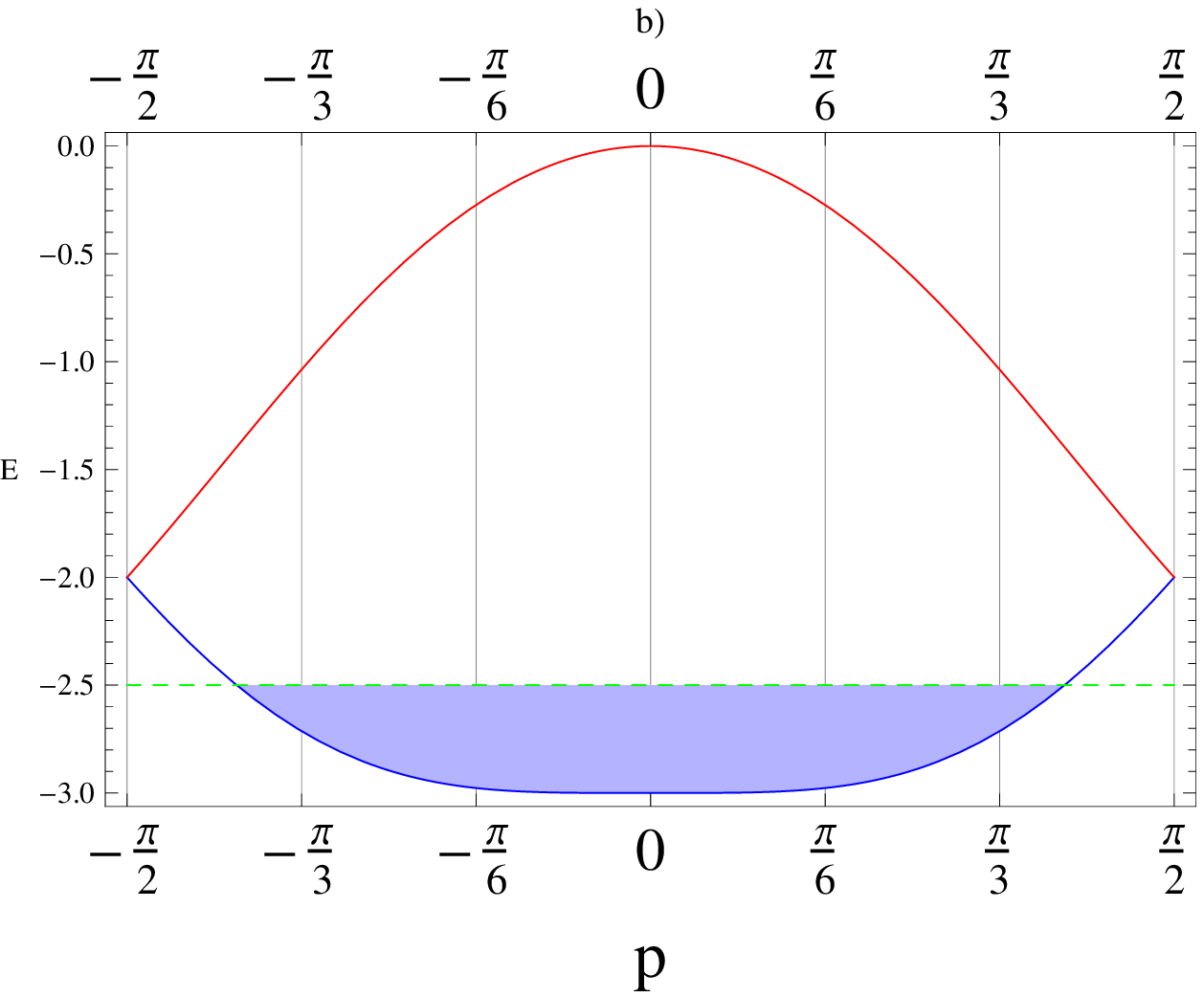} 
\end{center}
\end{minipage}
\begin{minipage}{0.32\linewidth}
\begin{center}
\includegraphics[width=\linewidth]{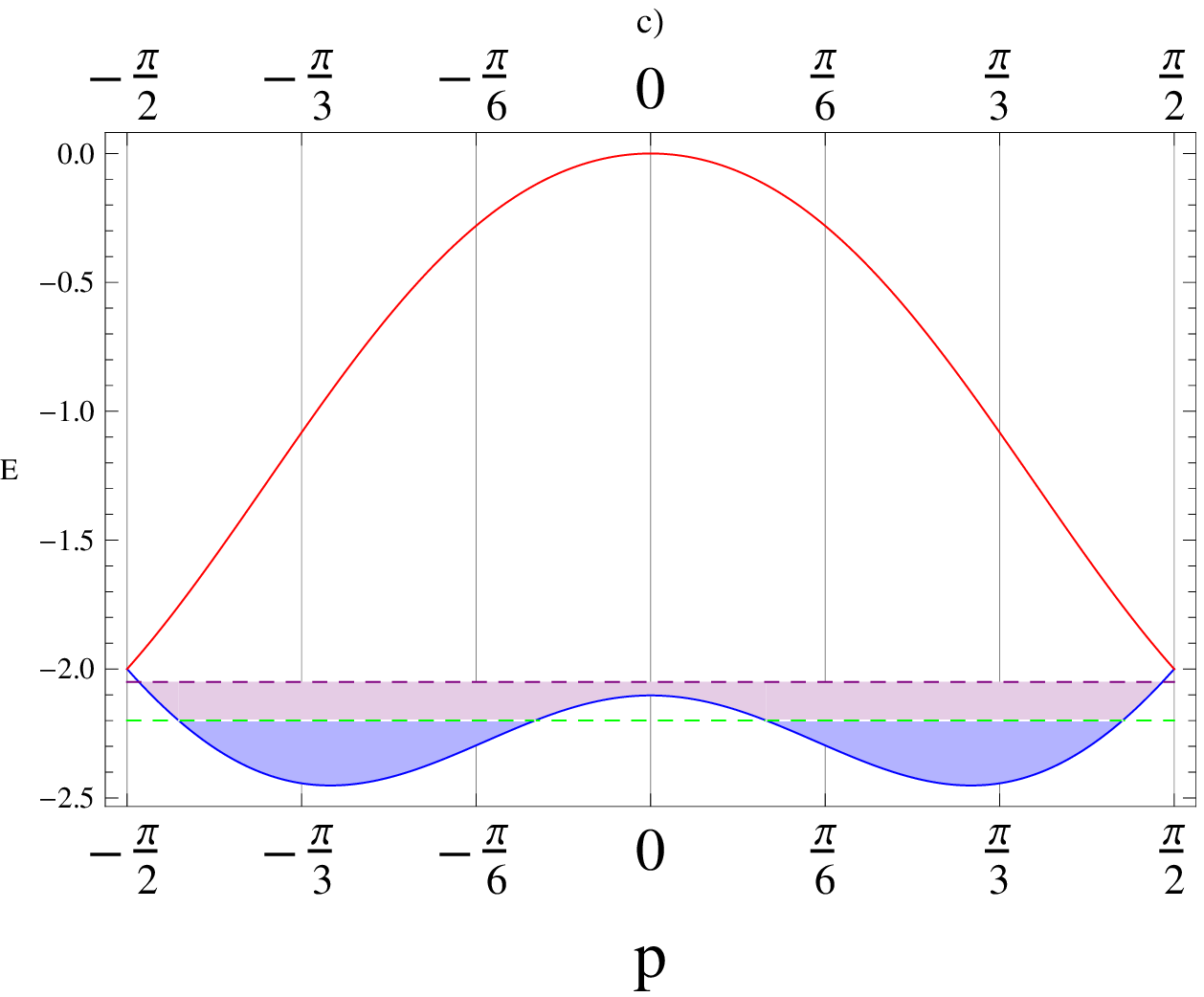} 
\end{center}
\end{minipage}
\caption{Dispersion Relation for spin/hole-waves: a) $\theta=0.2$, b) $\theta=\frac{\sqrt{3}}{3}$, c) $\theta=0.95$. The parity symmetry is spontaneously broken for $\theta > \frac{\sqrt{3}}{3}$. Holes in a background of up spins do not interact, therefore one may fill the one hole excitation spectrum up to zero density. This allow us to exactly study commensurate to incommensurate transition between ferromagnetic spin ordered phases.}
\label{figdisp}
\end{figure}

In \cite{FRAHM2}, it was thoroughly investigated the phase diagram of the ground-state at the half-filling limit (Figure \ref{fig01}a). By means of conformal invariance and finite-size scaling technique it was shown a discontinuity in the exponent related to the algebraic decay of spin-spin correlation function when crossing the transition line separating phases \textrm{III} and \textrm{VI}. Unfortunately, even with this evidence of a commensurate-incommensurate transition, it was only possible to calculate the corresponding transition line numerically.  By contrast, the line transition separating the ferromagnetic state from the others can be exactly calculated\cite{FRAHM2}. 
As a consequence of two-site translational invariance, one may propose two spin-wave excitations with momentum $p= \frac{2 \pi k}{ 2 L}( ~\text{mod }\pi) $, one for even sites and the other for odd sites. Therefore, for each momentum there will be a two-state module for Hamiltonian in the sector with one spin down. Diagonalizing the Hamiltonian in this sub-space provides the magnon dispersion
\eq
E^{\pm}(p)= \frac{1}{1+\theta^2} \left(-2+\theta^2(\cos{2 p}-1)\pm \sqrt{\theta^2 \sin^2{2 p}+4 \cos^2{p}}\right), \label{dispersion}
\en
also plotted in Figure \ref{figdisp}. At an external magnetic field slightly smaller than the critical value, the ferromagnetic state becomes unstable against the creation of these magnons and therefore the critical magnetic field is the minimum of the dispersion relation, since in our convention the ferromagnetic state has zero energy.

The dispersion relation undergoes a qualitative change by varying theta, similarly to the driving terms of above the NLIE. For theta smaller than $\frac{\sqrt{3}}{3}$ the minimum value occurs at $p=0$, whereas for theta larger than $\frac{\sqrt{3}}{3}$ there are two degenerate minima at momenta $p=\pm p_F$, reflecting a spontaneous breaking of parity symmetry.
To be more precise, one could form linear combinations of these two degenerate spin waves to get eigenstates of parity operator with opposite eigenvalues $\pm1$. It can be readily seen that the dispersion relation is quadratic everywhere at the transition line, except for the tri-critical point where it becomes quartic.

The second special limit $\theta \rightarrow 0$ is shown in Figure \ref{fig01}b. In this limit we have the usual t-J model which is a single chain without terms explicitly breaking $PT$-symmetriy. This case was firstly studied in \cite{KLUMPER-TJ} where thermodynamical properties were calculated at zero magnetic field. By varying the external magnetic field and particle density, we show that the model has four different phases, \textrm{I}, \textrm{III} and \textrm{VII},  \textrm{IX} and an additional ``trivial'' phase \textrm{X} (zero density $n\rightarrow 0$). There is no commensurate to incommensurate phase transition in this case and the phases \textrm{I} and \textrm{VII} are both metallic(Luttinger liquid) differing from each other by magnetic behaviour. Phase \textrm{I} is anti-ferromagnetic and gapless, while \textrm{VII} is ferromagnetic and gapped for spin excitations. Phases \textrm{III} and \textrm{IX} have already appeared in the previous limit. They are the insulating analogue of \textrm{I} 
and \textrm{VII}.

The behavior of ground-state correlation functions at long-wavelength was calculated from finite-size spectrum in \cite{KAWAKAMI}, while non-linear integral equations describing thermodynamics at finite temperatures have been provided in \cite{KLUMPER-TJ}. Both results confirmed Luttinger-liquid picture of phase \textrm{I}. For instance, calculation of (equal time) electron Green function at zero magnetic field and temperature gives a momentum distribution with algebraic singularity at Fermi momentum $\langle n_{p}\rangle=\langle n_{p_F}\rangle-c{|p-p_F|}^{\eta} \mbox{\text{sgn}}(p-p_F) $, where $\eta$ decreases monotonically from $\frac{1}{8}$ to $0$ as we move from half-filling to zero density. In \cite{KAWAKAMI} they also provided critical exponents for finite magnetic fields at densities near half-filling limit. However, both references did not fully study the phase diagram (Figure \ref{fig01}b), since the magnetic fields and densities were not appropriately chosen to achieve phase \textrm{VII}. From the 
fact that phase \textrm{VII} is adjacent to \textrm{IX}, we should have Luttinger-Liquid behaviour with frozen spin degree of freedom (see discussion below). This is because the magnetic field is already strong enough to order the system into a ferromagnetic state, as it could be observed in the magnetization profile.

In order to further describe the phase diagram of (\ref{Htjgen}) we have chosen to show the entropy against $\theta$ and $n$ to four different $H$ values: $H=0$, $H=0.25$, $H=3.0$, $H=4.5$. One may see from diagrams of Figures \ref{fig01}a and \ref{fig01}b that these four slices should encompass all possible phases.

\begin{figure}[t!]
\begin{minipage}{0.5\linewidth}
\begin{center}
\includegraphics[width=\linewidth]{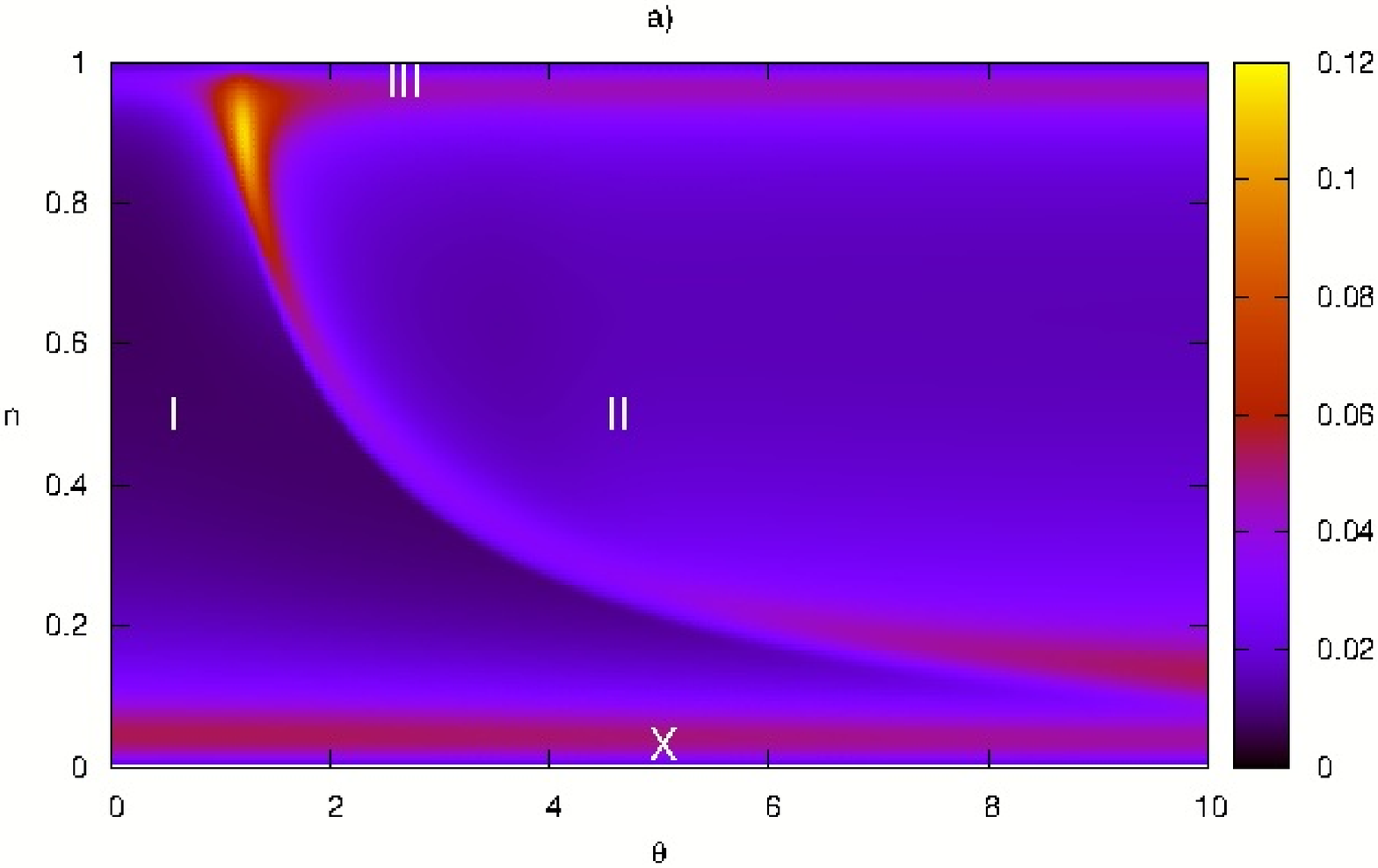} 
\end{center}
\end{minipage}%
\begin{minipage}{0.5\linewidth}
\begin{center}
\includegraphics[width=\linewidth]{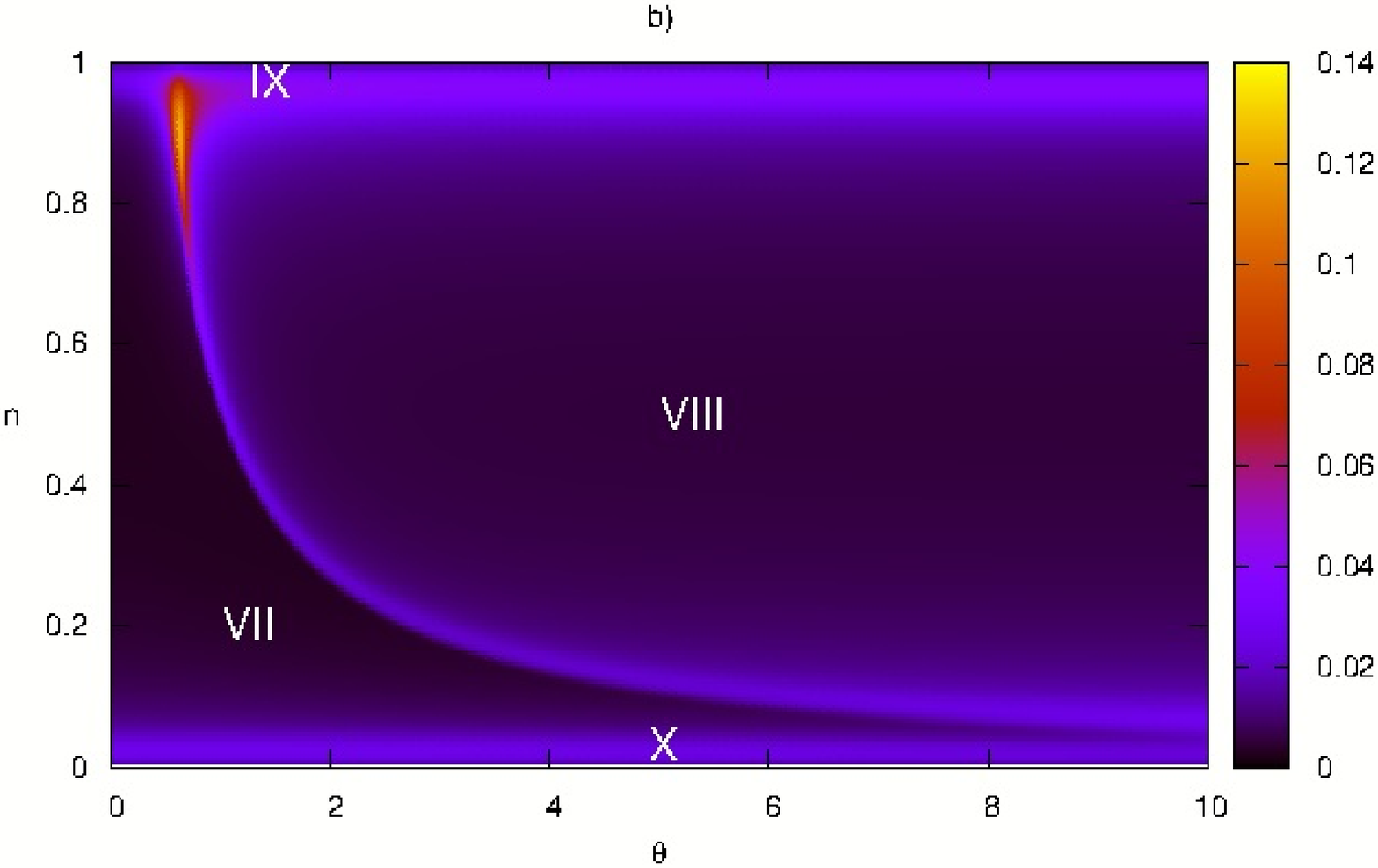} 
\end{center}
\end{minipage}
\caption{(Color online) Phase diagram obtained from entropy profile at $T=0.005$: a) $H=0$ and b) $H=4.5$.}
\label{fig1}
\end{figure}

For $H=0$, Figure \ref{fig1}a, we have an anti-ferromagnetic spin liquid. A closer look at Figure \ref{fig01}a reveals that no commensurate to incommensurate phase transition takes place at $n=1$, once the magnetic behaviour can only change in presence of magnetic field and the charge degree of freedom is frozen in this limit. Therefore, the line transition separating between phases \textrm{I} and \textrm{II} is of charge type and should end with an open point at $n=1$. Phase number \textrm{I} has been described above as a commensurate metallic commensurate anti-ferromagnetic phase\cite{KLUMPER-TJ}. Phase \textrm{II} is incommensurate metallic commensurate anti-ferromagnetic, since commensurate to incommensurate spin transition can only occur in presence of external magnetic field, see Figure \ref{fig01}a. The above distinction between \textrm{I} and \textrm{II} is also in agreement with the qualitative behaviour of auxiliary functions, Figures \ref{figI} and \ref{figII} in Appendix.  

For $H=4.5$, Figure \ref{fig1}b, we have that the magnetic order should be ferromagnetic in the whole $n-\theta$ plane. 
This is because the various anti-ferromagnetic interactions competing with ferromagnetic order induced by external field do only contribute when there are particles populating the chain. Therefore, such interactions exerts its full influence at half filling limit. Since at half filling any external field larger than 4 is sufficient to drive the ground-state to ferromagnetic ordering, the same is true for any particle density. Analogously to the $H=0$ value, no commensurate-to-incommensurate phase transition takes place at half-filling, therefore the line separating between phases \textrm{VII} and \textrm{VIII} is of charge type and terminates with an open point at $n=1$. Phase \textrm{VII} and \textrm{IX} are described as commensurate metallic ferromagnetic phase and insulating ferromagnetic phase, respectively. Phase \textrm{VIII} is the incommensurate metallic ferromagnetic phase.

Similar to the spin-waves proposed to describe the line transition separating the ferromagnetic state in  Figure \ref{fig01}a, one may propose ``hole-waves'' to describe the line transition separating the insulating phase in Figure \ref{fig1}b. The dispersion relation for one hole excitation is (\ref{dispersion}) again. From this analysis, one could derive the critical chemical potential which drives the system to the half-filling limit. However, we find a major difference between hole-waves and spin-waves, namely, excitations with more holes now can be readily calculated. This is due to the fact that holes in the background of up spins do not interact, which can also be seen from Bethe ansatz equations for row-to-row transfer matrix
\begin{align}
{\left[ \frac{a_+(\lambda_r^1) a_+(\lambda_r^1-\im \theta)}{b(\lambda_r^1) b(\lambda_r^1-\im \theta)}\right]}^L &=\left[\prod_{\stackrel{k=1}{k \neq r}}^{m_1}\frac{a_-(\lambda_r^1-\lambda_k^1) b(\lambda_k^1-\lambda_r^1)}{b(\lambda_r^1-\lambda_k^1) a_+(\lambda_k^1-\lambda_r^1)}\right] \prod_{k=1}^{m_2} \frac{b(\lambda_k^2-\lambda_r^1)}{a_-(\lambda_k^2-\lambda_r^1)}, \nonumber
\\
\prod_{k=1}^{m_1} \frac{a_-(\lambda_r^2-\lambda_k^1)}{b(\lambda_r^2-\lambda_k^1)}&=\prod_{\stackrel{k=1}{k\neq r}}^{m_2} \frac{a_+(\lambda_r^2-\lambda_k^2) b(\lambda_k^2-\lambda_r^2)}{b(\lambda_r^2-\lambda_k^2) a_-(\lambda_k^2-\lambda_r^2)}.
\end{align}
In sector $m_1=2L-N$, $m_2=0$, the second set of equations becomes trivial while the first one does not possess scattering terms. Therefore, this set of sectors resembles the free-fermion with exclusion principle being implemented by inequality of Bethe ansatz roots. Hence, the ground-state is found by filling the lowest energy states. In this way, the metallic phases of Figure \ref{fig1}b are actually a non-interacting limit of Luttinger liquid, the free-fermions.

This property allows us to determine the remaining transition line in Figure \ref{fig1}b, which separates phases \textrm{VII} and \textrm{VIII}. At theta larger than $\frac{\sqrt{3}}{3}$, one may change from commensurate to incommensurate phase by varying the particle density. The transition line is obtained by setting the Fermi level at the local maximum of dispersion relation (\ref{dispersion}), see Figure \ref{figdisp}. Let $h$ be this Fermi level, then the parametric equation of transition line is given by
\eq
(\theta,n)=\left(\sqrt{-\left(1+\frac{4}{h}\right)},\begin{cases}
\frac{\arccos\left(\frac{4+7 h+2 h^2}{4+h}\right)}{2 \pi} &\text{if}~-2\leq h\leq0\\
1-\frac{\arccos\left(\frac{4+7 h+2 h^2}{4+h}\right)}{2 \pi}  &\text{if}~-3\leq h\leq-2\end{cases}
\right),
\en
which is in good agreement with numerical findings at finite temperature ($T=0.005$) from the non-linear integral equations.

The simplicity of free-fermion also allows us to exactly calculate some correlation functions at any distance. For instance, let us consider the correlation $G(j_2-j_1)=\langle c_{j_1 \uparrow} c_{j_2 \uparrow}^{\dag} \rangle$ for $j_1$ and $j_2$ odd. In thermodynamic limit we find
\eq
G(x)=\frac{1}{\pi} \int_{E^+(p)\leq E_F}\frac{{\rm e}^{-\im p x}}{1+{S^+}^2(p)} {\rm d}p+\frac{1}{\pi} \int_{E^-(p)\leq E_F}\frac{{\rm e}^{-\im p x}}{1+{{S^-}^2(p)}} {\rm d}p,
\en
where $S^{\pm}(p)=\frac{2 \cos(p)}{\theta \sin{2p} \pm\sqrt{\theta^2 \sin^2{2 p}+4 \cos^2{p}}}$. Additionally, let us limit ourselves to $\theta$ smaller than $1$ and $n$ larger than $\frac{1}{2}$. In this way, we may completely ignore the upper spectrum $E^+(p)$ while we still can observe qualitative differences when changing from \textrm{VII} to \textrm{VIII}. We find
\eq
\Re\{G(x)\}=\begin{cases}
          \frac{2}{\pi x}\sin{p_F x}&\text{in phase \textrm{VII}}\\
          \frac{4 }{\pi x} \sin(({p_F}_2-{p_F}_1) \frac{x}{2}) \cos(({p_F}_1+{p_F}_2)\frac{x}{2})&\text{in phase \textrm{VIII}}
          \end{cases},
\en
where ${p_F}$ is the sole positive Fermi momentum in the commensurate phase, while ${p_F}_1$ and ${p_F}_2$ are the two positive Fermi momenta in the incommensurate phase. As can be expected, in the thermodynamic limit ${p_F}_2-{p_F}_1$ is not commensurate with ${p_F}_2+{p_F}_1$, except for infinitely many countable values of $n$. Therefore, the system is in an incommensurate floating phase\cite{BAK}, and may exhibit modulated oscillations with no definite period.

\begin{figure}[t!]
\begin{minipage}{0.5\linewidth}
\begin{center}
\includegraphics[width=\linewidth]{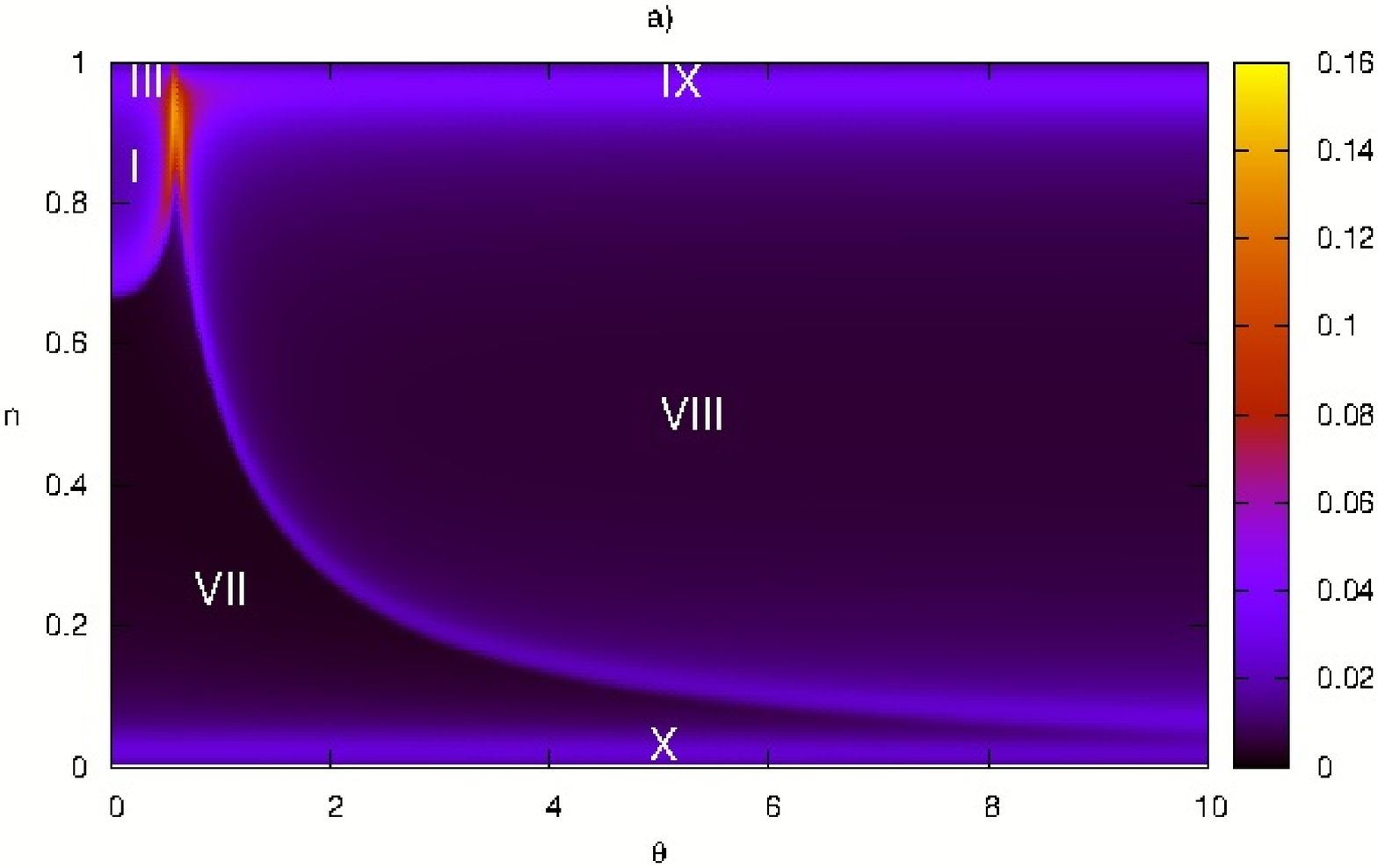} 
\end{center}
\end{minipage}%
\begin{minipage}{0.5\linewidth}
\begin{center}
\includegraphics[width=\linewidth]{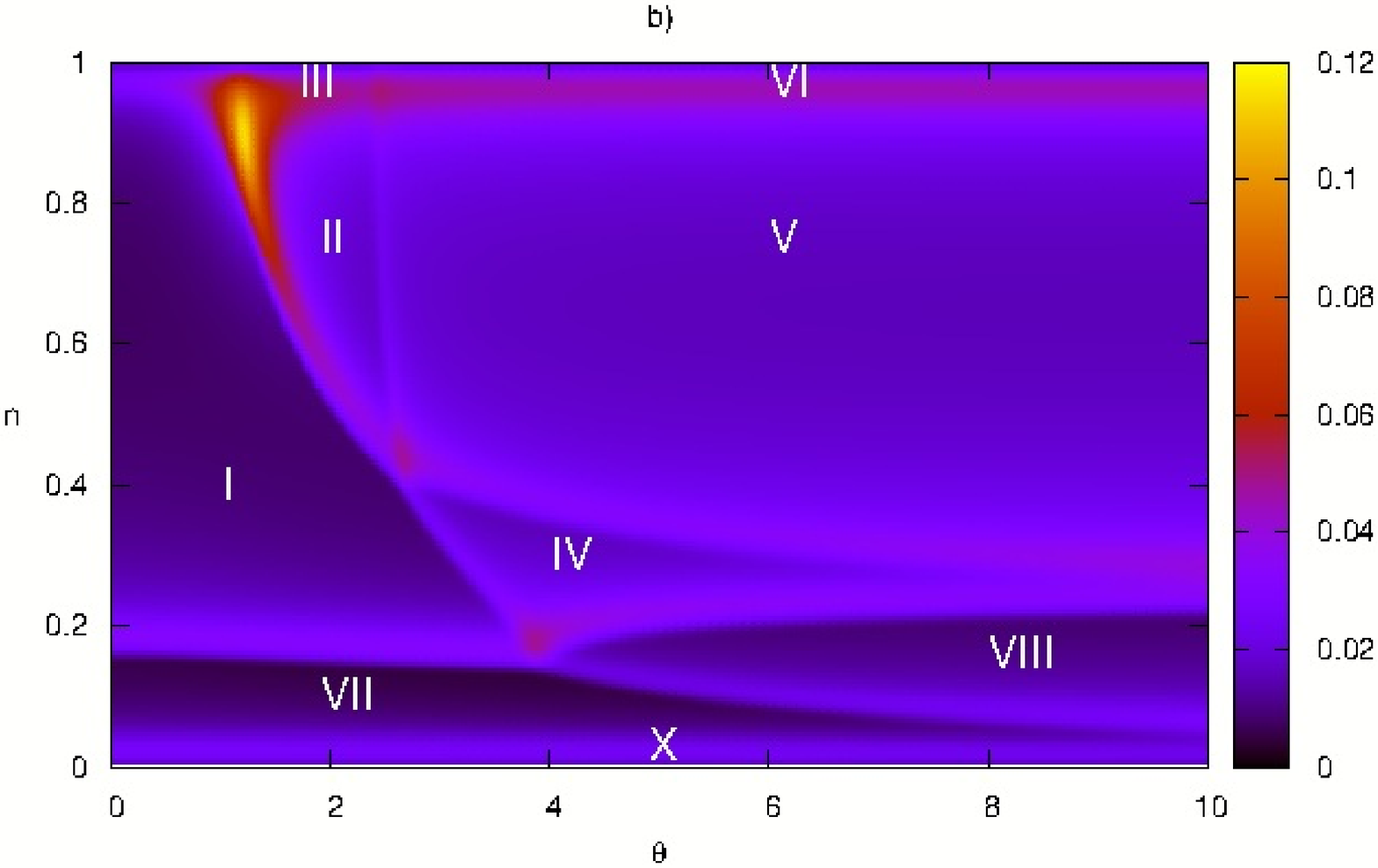} 
\end{center}
\end{minipage}
\caption{(Color online) Phase diagram obtained from entropy profile at $T=0.005$: a) $H=3$ and b) $H=0.25$.}
\label{fig3}
\end{figure}

Next we examined the case $H=3$, Figure \ref{fig3}a. All phases in this diagram have already been described before. Besides showing the phase boundaries as we change the external magnetic field, there is an interesting feature at half-filling termination of transition lines. Here the transition ending point separating phases \textrm{I}, \textrm{VII} and \textrm{VIII} is not open as before. From \textrm{I} to \textrm{VII}, or \textrm{I} to \textrm{VIII} along $n=1$ there is a magnetic transition from commensurate anti-ferromagnetic behaviour to ferromagnetic behaviour. Although the decreasing line, which marks charge transition, is open at the ending point, the increasing line, related to magnetic transition, is closed in correspondence with Figure \ref{fig01}a.

The last case is for $H=0.25$, Figure \ref{fig3}b, which includes two new phases. In this case, we have a mixture of the spin and charge effects and therefore the situation is very complicated. There is no simple description as given before. However, one may expect based on the behaviour of the auxiliary functions that the phase \textrm{IV} is a commensurate metallic incommensurate anti-ferromagnetic phase. It could only exists because of a ``crossing'' of a charge line transition and a spin line transition. As consequence, phase \textrm{II} becomes limited, while there arises another unlimited phase, \textrm{V}. This last phase express charge and spin incommensurability, therefore the incommensurate metallic incommensurate anti-ferromagnetic phase.

As it has been mentioned, to all ten phases there corresponds a qualitative behaviour of auxiliary functions. We show this in Appendix, where auxiliary functions for typical points of each phase in Figures \ref{fig1}  and \ref{fig3} may be compared.

In all phase diagrams presented here we have chosen to study phase transitions as a function of $n$ instead of $\mu$. Nevertheless, we can follow \cite{TJHIGHER} and consider  $\mu$ as the external voltage. As it happens, exactly at $T=0$, it is not necessary to take $\mu \rightarrow \infty$ to recover the half-filling limit. Comparing the NLIE, or looking at auxiliary functions behaviour, we see that the condition is the vanishing of $\frac{\ln \mathfrak{C}}{\beta}$ or, equivalently, $\frac{\ln \mathfrak{c}}{\beta}\leq0$. Therefore, we find 
\eq
\mu'={\max_x}\left[F_{\theta}(x)+\lim_{\beta \rightarrow \infty}\frac{K}{\beta}\ast (\ln \mathfrak{B}\bar{\mathfrak{B}})(x)\right],
\en
where $\mathfrak{B}$ and $\bar{\mathfrak{B}}$ are determined from NLIE for the Heisenberg multi-chain\cite{TAVARES}. For zero magnetic field and $\theta$ we get $\mu'=2 \ln 2$, in accordance to \cite{TJHIGHER}. Similarly, the limit $n \rightarrow 0$ requires the vanishing of $\frac{\ln\mathfrak{B}}{\beta}$ and $\frac{\ln\bar{\mathfrak{B}}}{\beta}$ for any $H$ and $\frac{\ln\mathfrak{C}}{\beta}$ may be approximated to $\frac{\ln \mathfrak{c}}{\beta}$. Therefore we find for the trivial phase $\mu'=-\frac{H}{2}$. Following the arguments of \cite{TJHIGHER}, if one starts with a fixed $\mu$ such that $n=1$ for given $\theta$, by varying $\theta$ one can get $n<1$ if the new $\theta$ has a larger $\mu(\theta)$ corresponding to $n=1$. Therefore we would have the appearance of holons without changing applied voltage. Nevertheless, since $\mu$ is always negative in absence of magnetic field, we cannot say it is a spontaneous charge ordering. First-order transitions are related to dressed energies having maxima points 
larger than the
asymptotic limit value at $x \rightarrow \infty$. As we can see in Appendix, there is no solution for either $\frac{\ln \mathfrak{b}}{\beta}$ or $\frac{\ln \mathfrak{c}}{\beta}$(inverted Dirac sea) with such behaviour.

\section{Conclusion}
\label{conclusion}

In this paper we applied the quantum transfer matrix approach to the case of an integrable generalization of the super-symmetric t-J model containing interactions explicitly breaking $PT$-symmetry. We derived a new set of non-linear integral equations for the thermodynamical properties of the generalized t-J model.

We solved the non-linear integral equations as a function of temperature, magnetic field and chemical potential/density of particles. This reveals us a rich $n-\theta-H$ ground state phase diagram with ten different phases. We used the previous knowledge of the phase diagram at two special planes ($n-H$ at $\theta=0$ where the model reduces to the usual t-J model and $H-\theta$ at $n=1$ which reduces to the Heisenberg model with competing interaction) and the analysis of the auxiliary function in order to classify the different phases. Besides, we present the physical description in the half-filling limit as well as in the high-magnetic field limit. This gives us the possible scenarios for the ground state phase diagram, based on the computation of the physical properties at low but finite temperature. We have the combinations of insulating/metallic order with ferromagnetic/anti-ferromagnetic order of commensurate/incommensurate nature. These results show an interesting and rich phase diagram for this system 
of strongly correlated electrons. However the phases II and IV still needs further investigation, since their fully description is beyond our approach. Due to the number of new different phases, we hope that our exact results will be useful for understanding experimental results of some quasi one-dimensional systems.

We expect that our results could useful to the study of the phase diagram of similar systems. We also expect that the quantum transfer matrix can be generalized  to other spin chains invariant by super-algebras, e.g $Osp(1|2)$ case. We hope to address this problem in the future.

\section*{Acknowledgments}

The authors thank Andreas Kl\"{u}mper for useful discussions.
T.S. Tavares thanks FAPESP for financial support through the grant 2013/17338-4. G.A.P. Ribeiro acknowledges financial support through the grants 2015/01643-8 and 2015/07780-7, S\~ao Paulo Research Foundation (FAPESP).

\section*{Appendix: Qualitative differences of auxiliary functions for the different phases}

\setcounter{figure}{0}
\renewcommand{\thefigure}{A.\arabic{figure}}

In this appendix we would like to present the qualitative aspect of auxiliary functions $\frac{\ln b(x)}{\beta}$ and $\frac{\ln c(x)}{\beta}$ for each phase.

Solving equation (\ref{nlieq}) for given $T$, $H$ and $n(\mu,T,H)$ gives the auxiliary functions $ \frac{\ln \mathfrak{b}(x)}{\beta}$, $ \frac{\ln \bar{\mathfrak{b}}(x)}{\beta}$ and $ \frac{\ln \mathfrak{c}(x)}{\beta}$. In the limit $\beta \rightarrow \infty$, where quantum phase transitions takes place, these functions must show some difference in qualitative behaviour for different phases, reflecting the non-analyticity of thermodynamical properties at critical lines. Therefore, the auxiliary functions can be used to endorse our phase classification of Table \ref{table1}.

The presentation of auxiliary functions can be simplified by noting that for $H \geq 0$ one can drop $\bar{\mathfrak{b}}(x)$. This is because, for $H=0$, functions $\mathfrak{b}(x)$ and $\bar{\mathfrak{b}}(x)$ are complex conjugate of each other, while for $H>0$,  $\bar{\mathfrak{b}}(x)$ vanishes in the limit $\beta \rightarrow \infty$. Besides, the imaginary part of $ \frac{\ln \mathfrak{b}(x)}{\beta}$ and $ \frac{\ln \mathfrak{c}(x)}{\beta}$ revealed to be very small. Therefore, we shall restrict to the study of real part of functions $ \frac{\ln \mathfrak{b}(x)}{\beta}$ and $ \frac{\ln \mathfrak{c}(x)}{\beta}$.

\begin{figure}[t!]
\begin{center}
\begin{minipage}{0.35\linewidth}
\includegraphics[width=\linewidth]{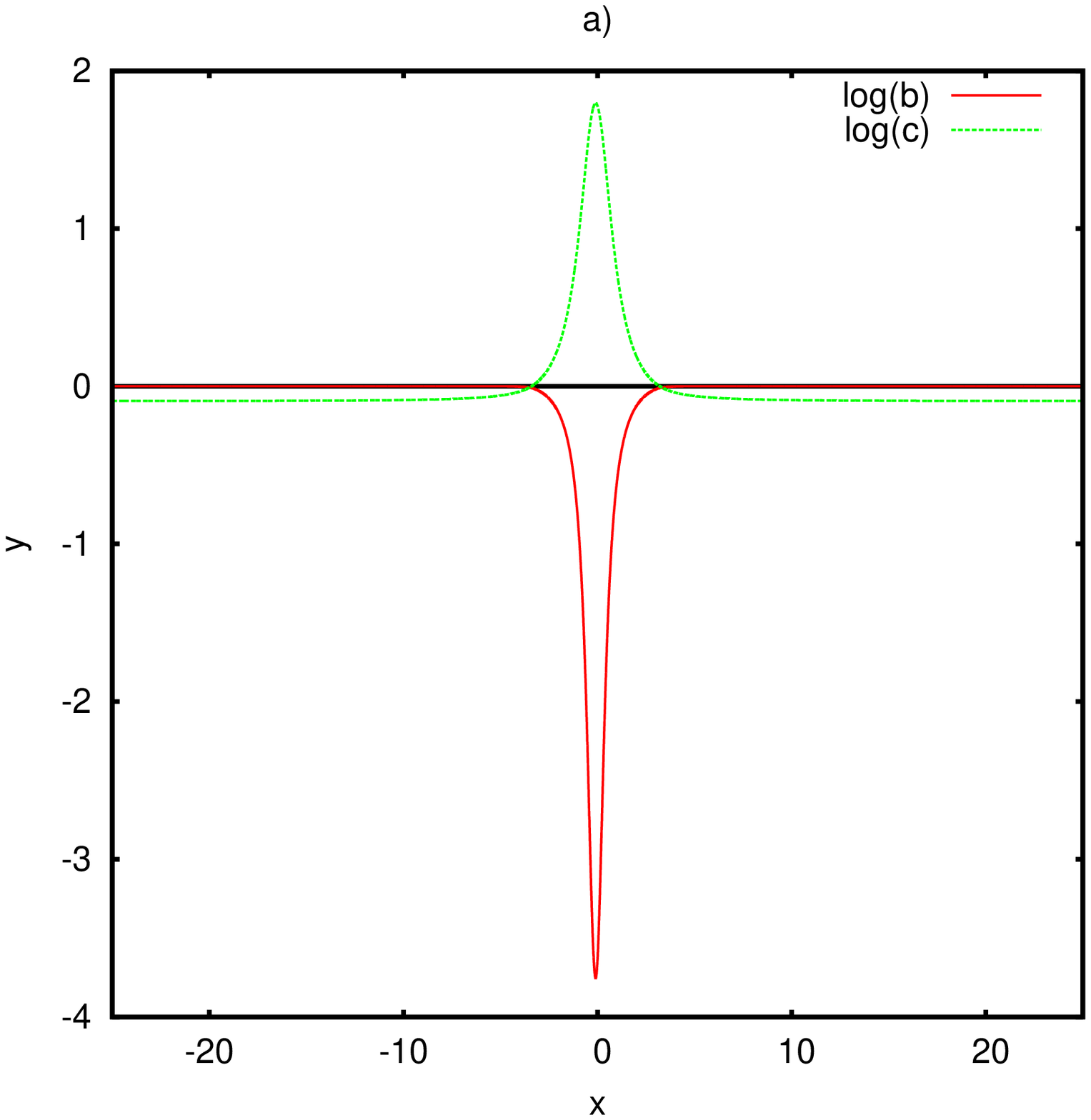}
\includegraphics[width=\linewidth]{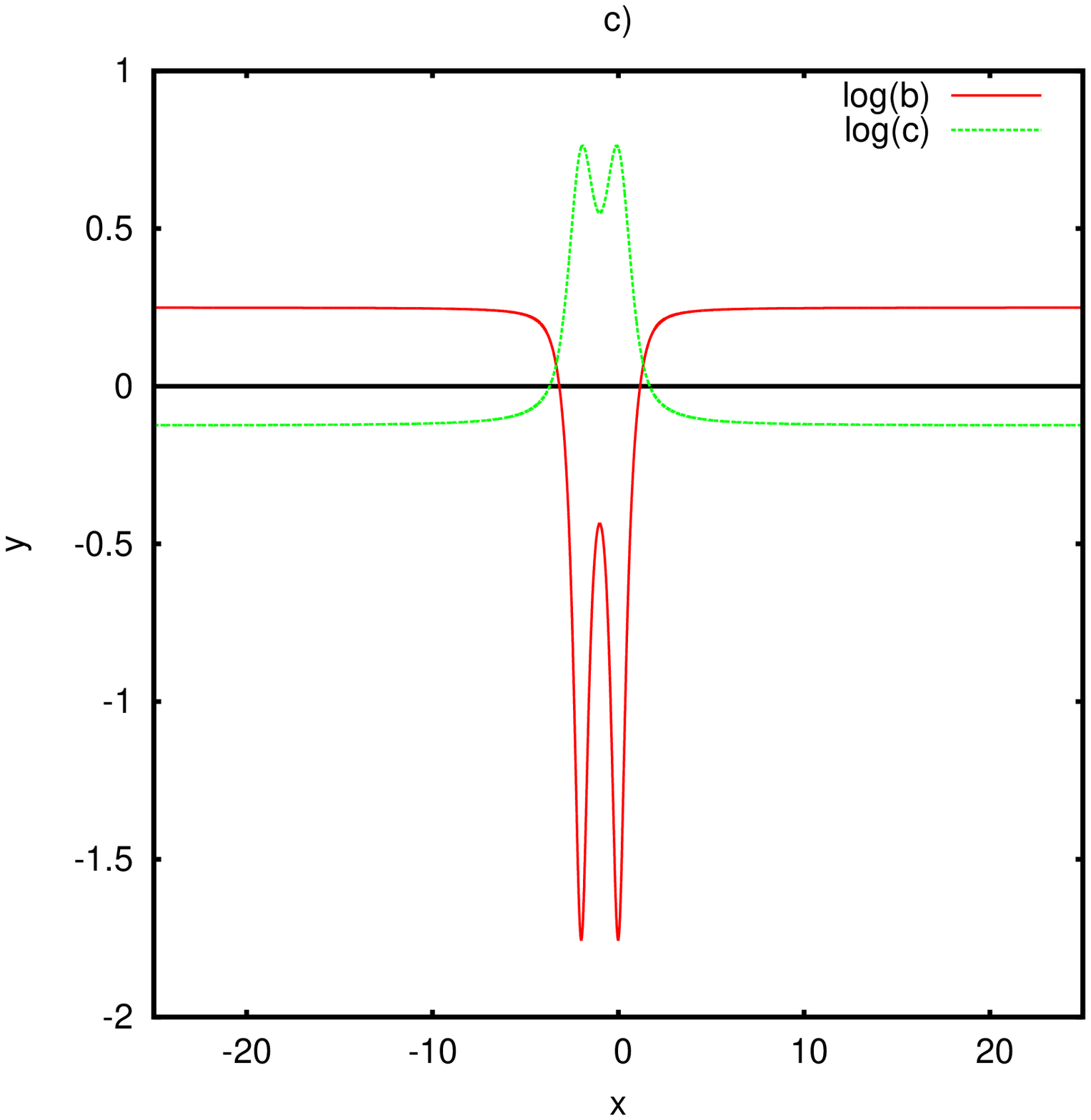}
\end{minipage}%
\begin{minipage}{0.35\linewidth}
\includegraphics[width=\linewidth]{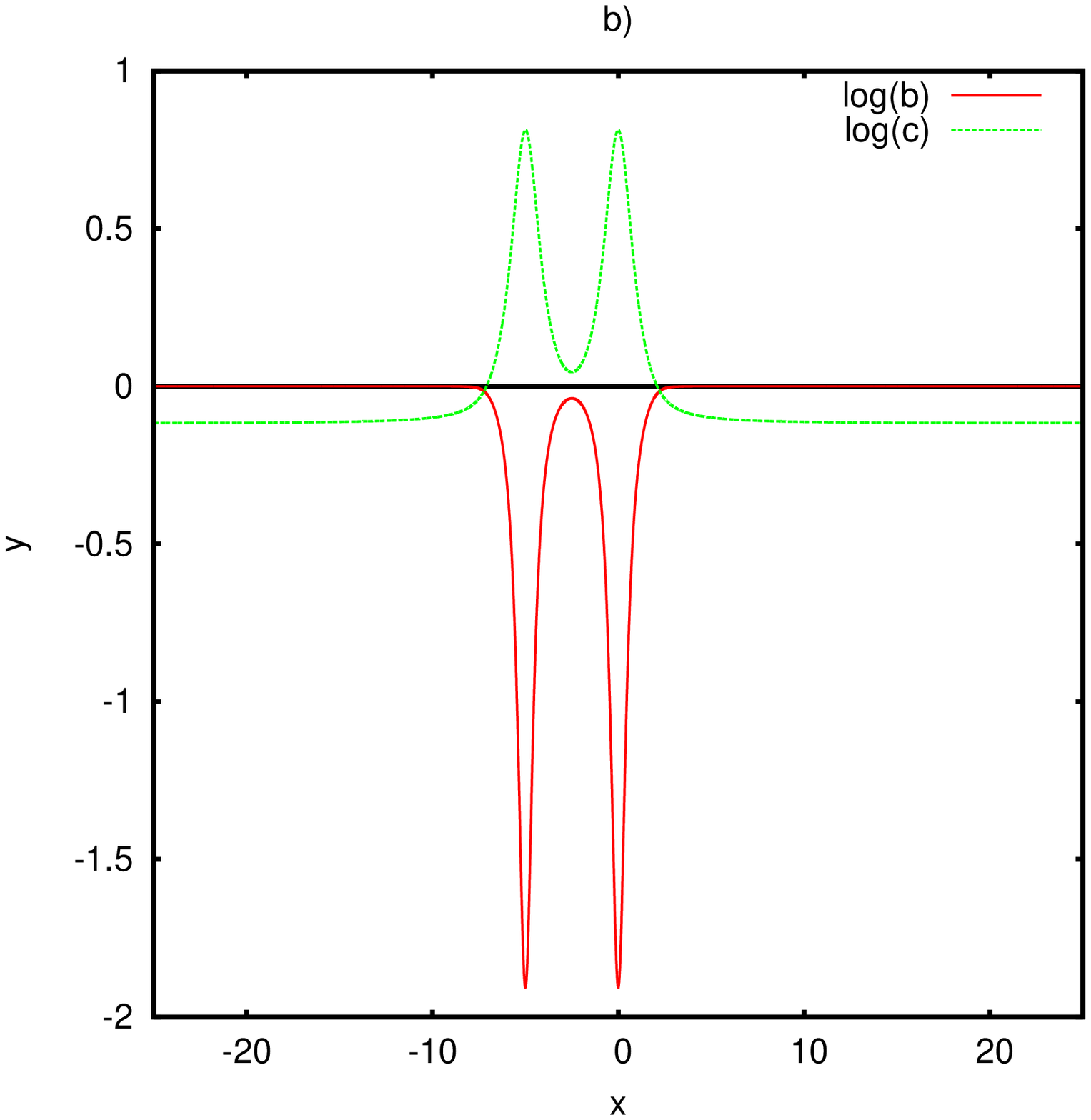}
\includegraphics[width=\linewidth]{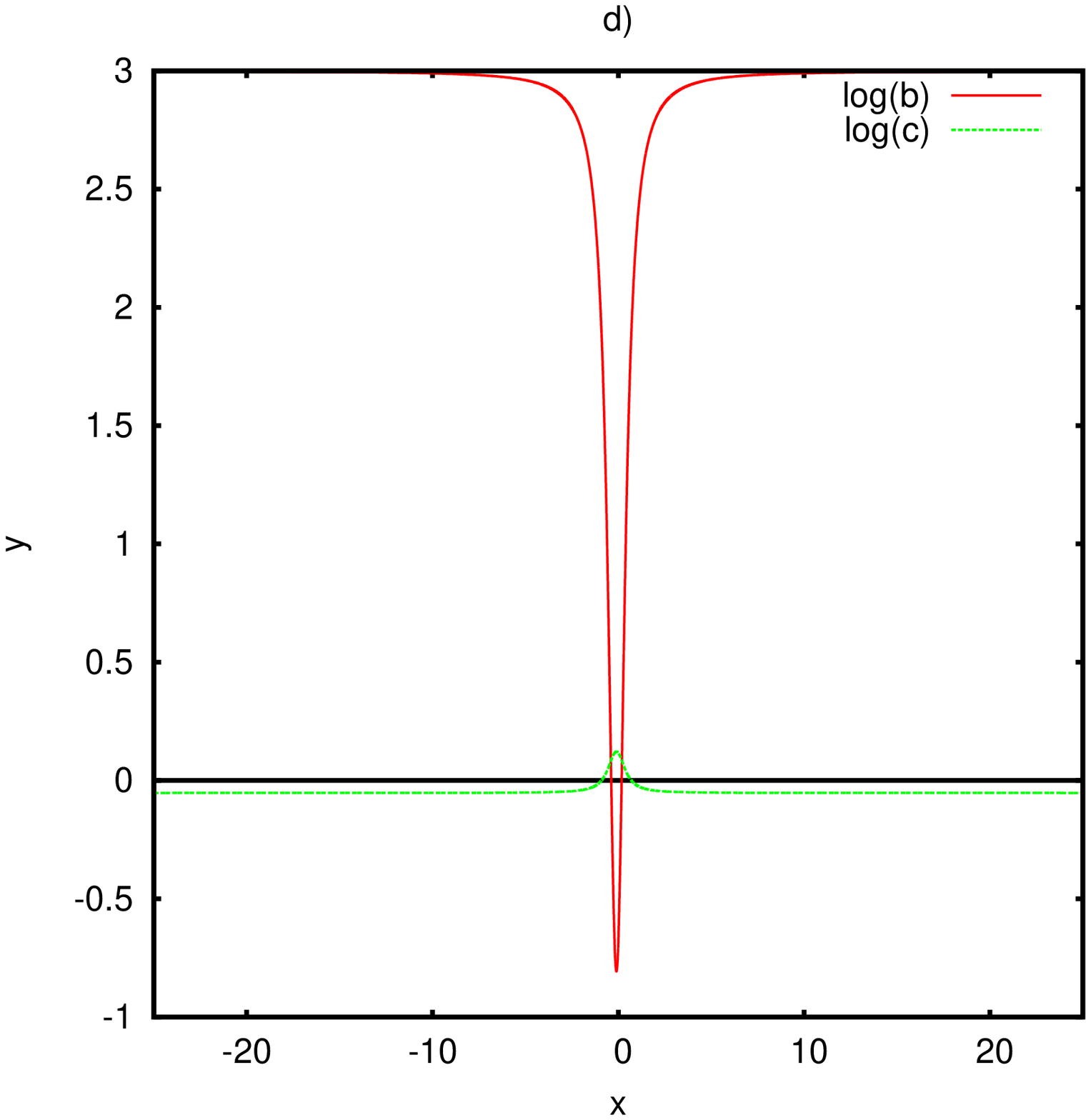}
\end{minipage}
\end{center}
\caption{(Color online) Phase I: Commensurate metallic commensurate anti-ferromagnetic: $\frac{\ln b}{\beta}$ has one dirac sea and $\frac{\ln c}{\beta}$ has one inverted dirac sea. a) $H=0,~\theta=0.2,~n=0.2$, b) $H=0,~\theta=5,~n=0.2$, c) $H=0.25,~\theta=2,~n=0.3$ d) $H=3,~\theta=0.2,~n=0.8$. }
\label{figI}
\end{figure}

\begin{figure}[t!]
\begin{center}
\begin{minipage}{0.35\linewidth}
\includegraphics[width=\linewidth]{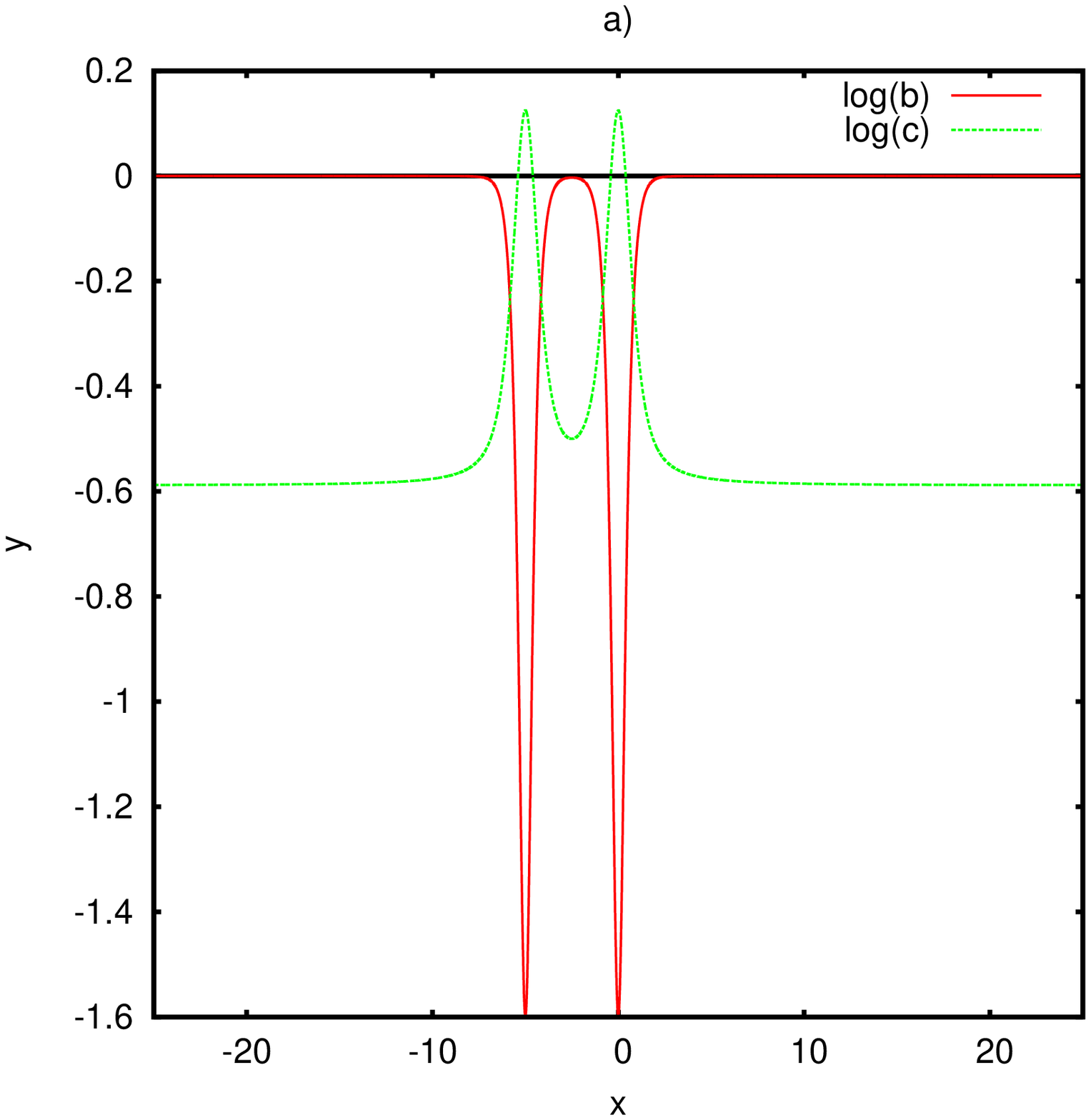}
\end{minipage}
\begin{minipage}{0.35\linewidth}
\includegraphics[width=\linewidth]{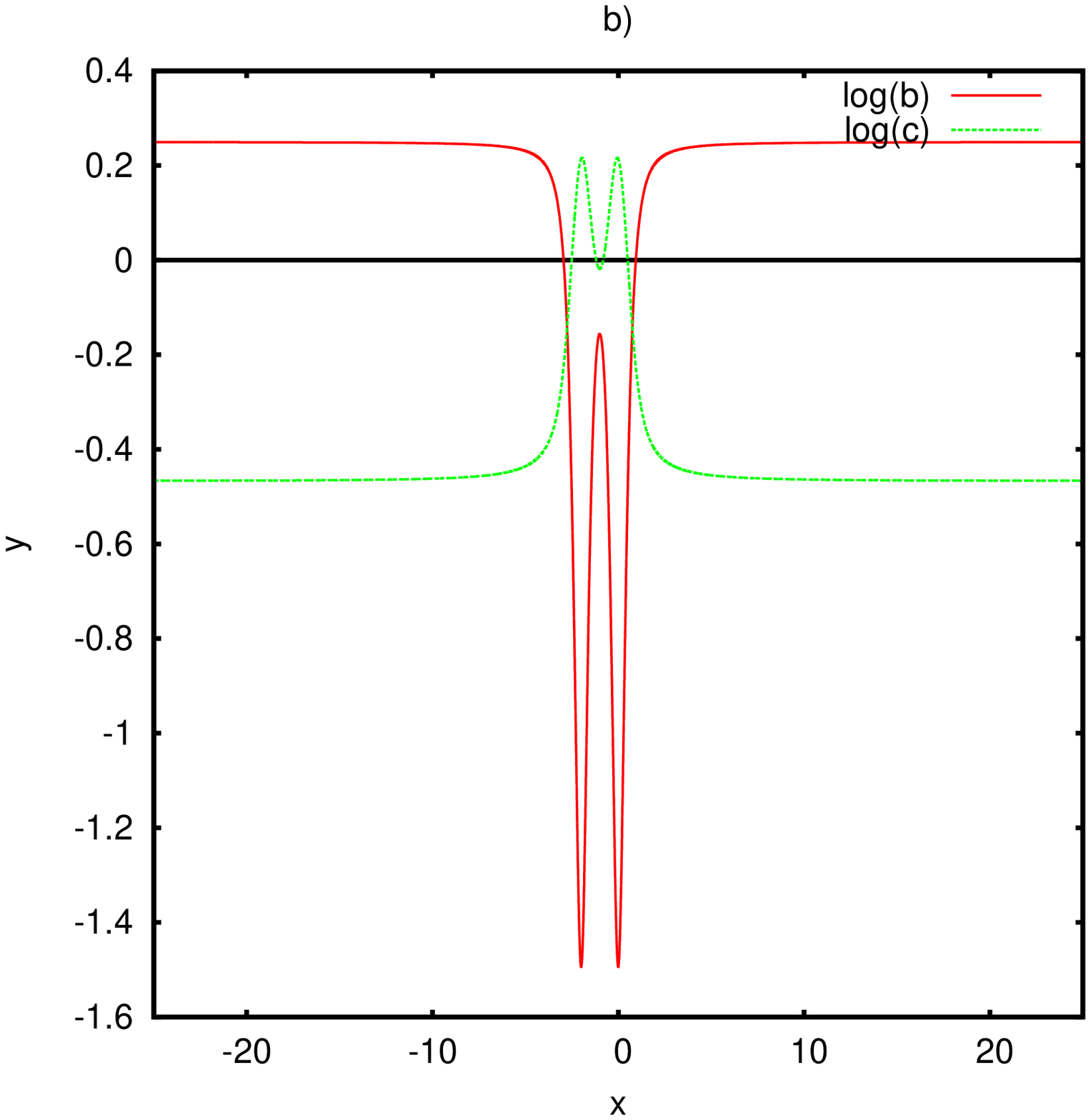}
\end{minipage}
\end{center}
\caption{(Color online) Phase II: Incommensurate metallic commensurate anti-ferromagnetic: $\frac{\ln b}{\beta}$ has one dirac sea and $\frac{\ln c}{\beta}$ has two inverted dirac sea. a) $H=0,~\theta=5,~n=0.8$, b) $H=0.25,~\theta=2,~n=0.6$. }
\label{figII}
\end{figure}

\begin{figure}[htb]
\begin{center}
\begin{minipage}{0.32\linewidth}
\includegraphics[width=\linewidth]{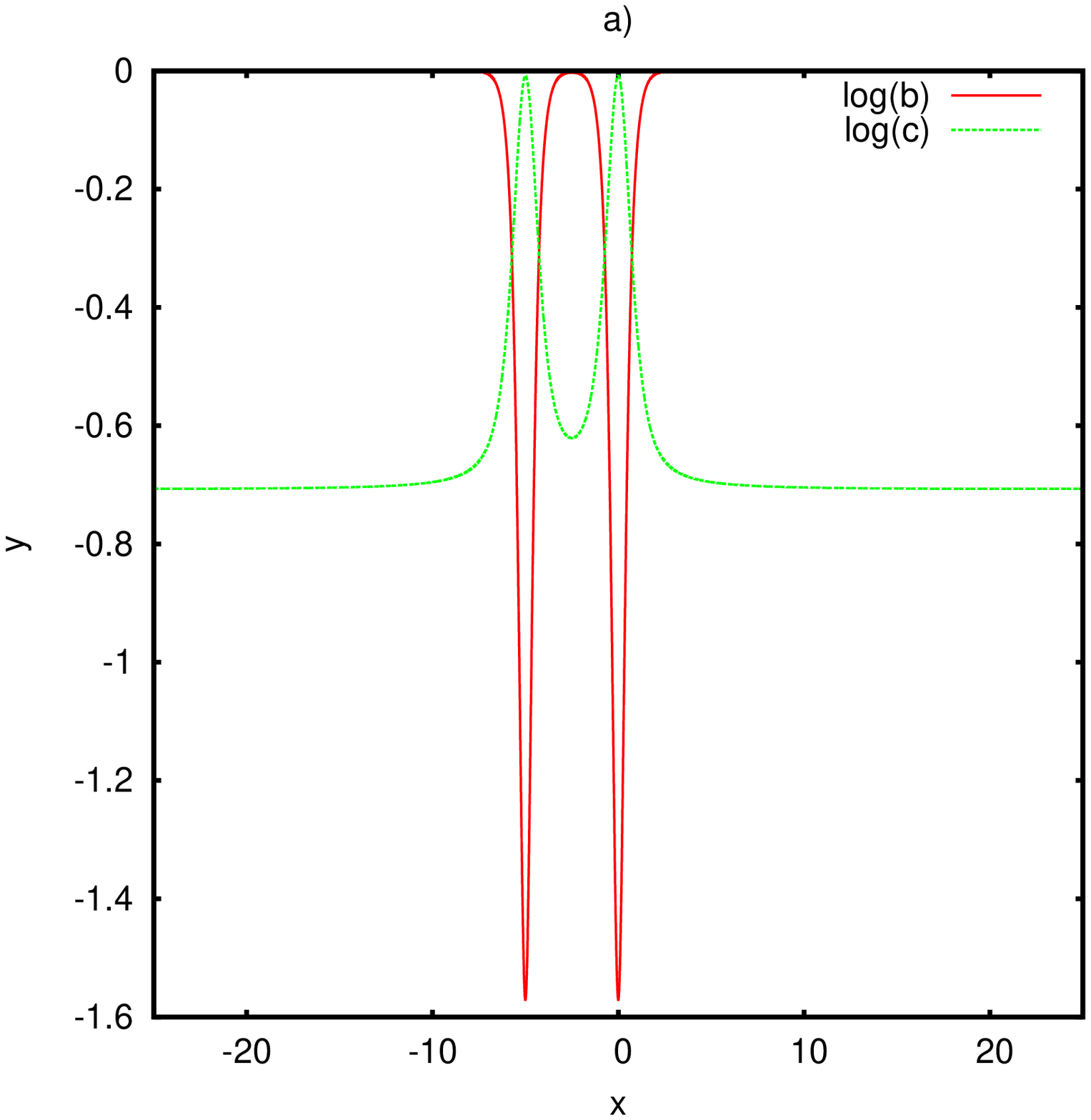}
\end{minipage}
\begin{minipage}{0.32\linewidth}
\includegraphics[width=\linewidth]{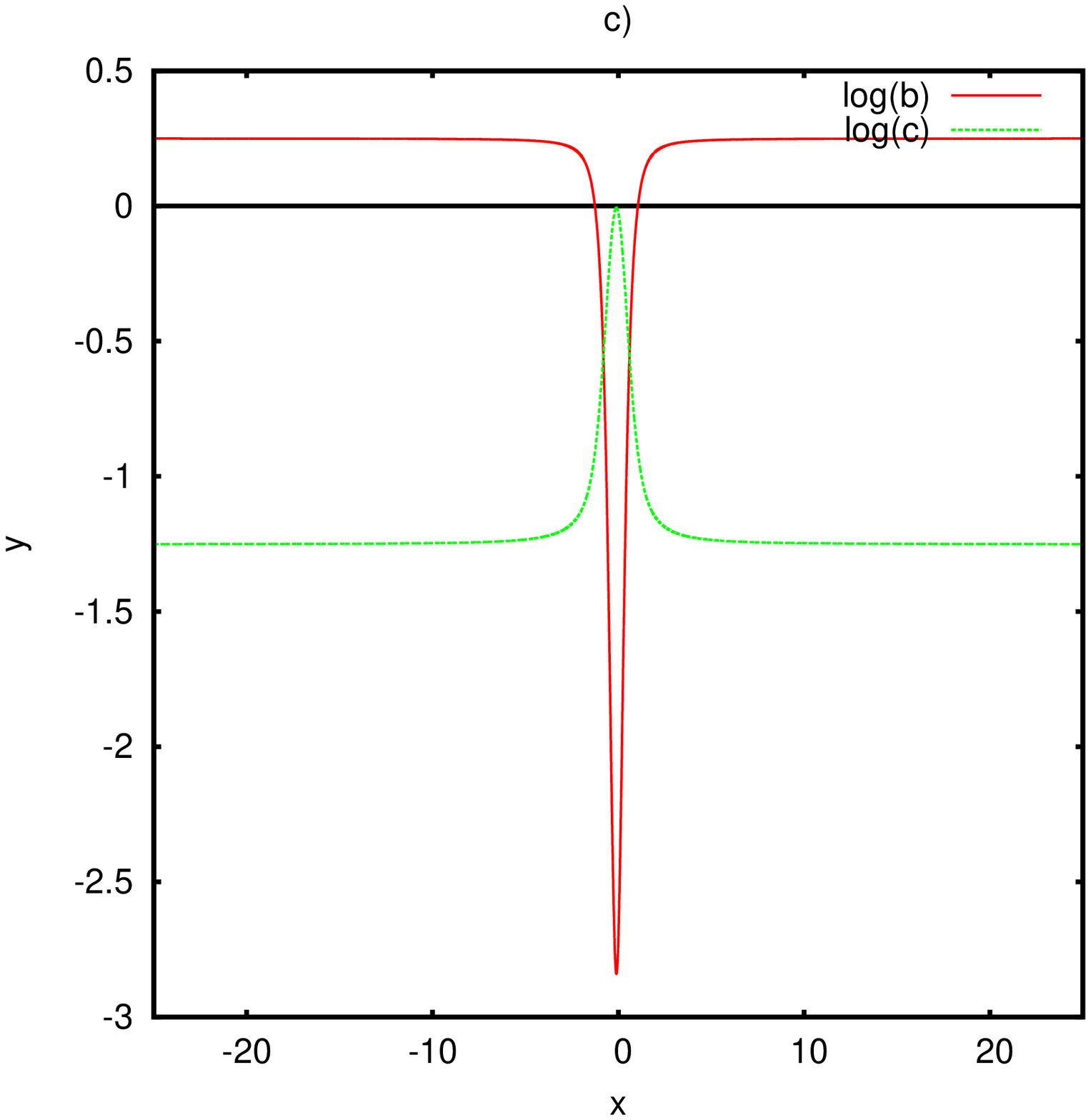}
\end{minipage}
\begin{minipage}{0.32\linewidth}
\includegraphics[width=\linewidth]{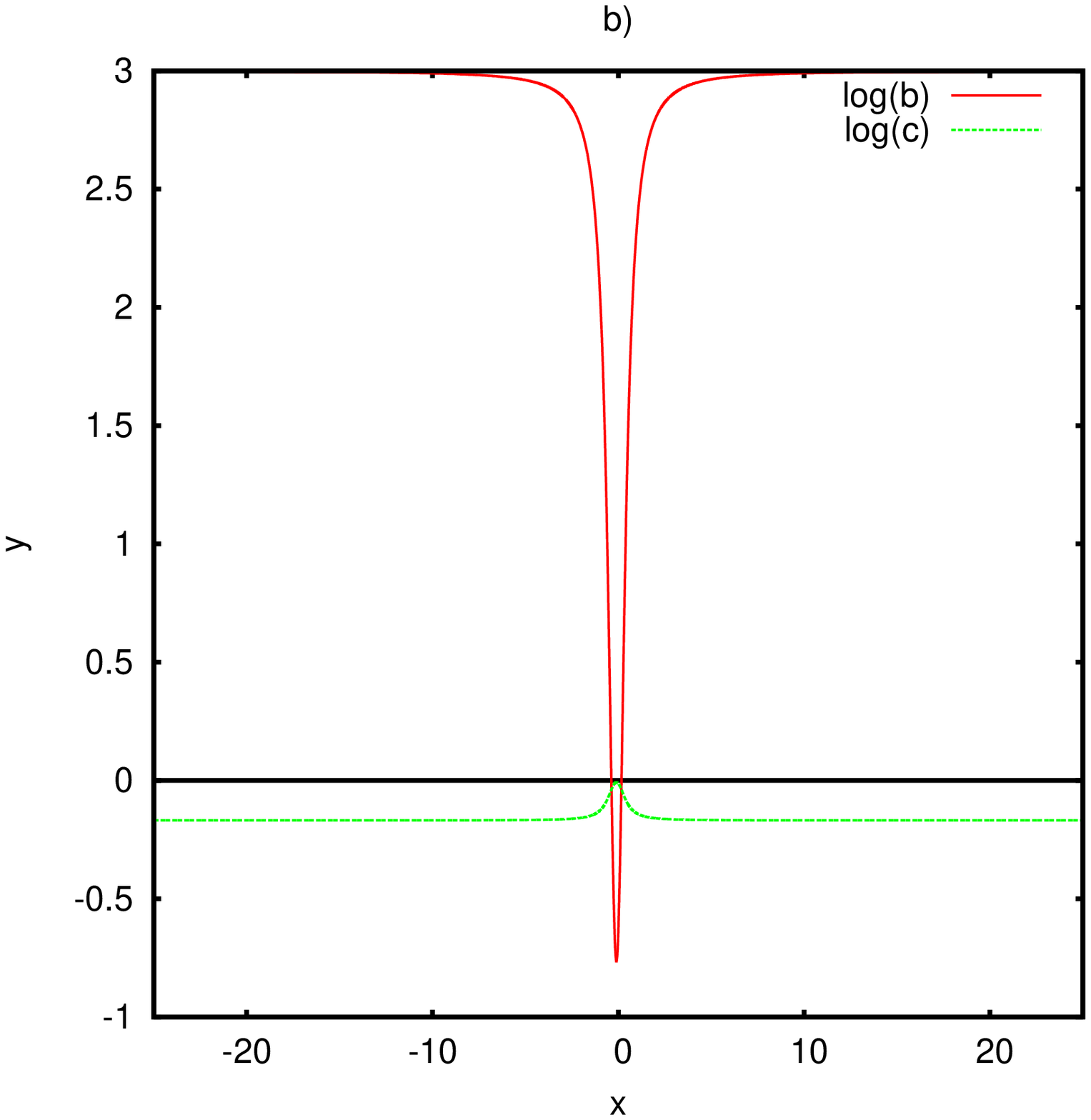}
\end{minipage}
\end{center}
\caption{(Color online) Phase III: Insulating commensurate anti-ferromagnetic:  $\frac{\ln b}{\beta}$ has one dirac sea and $\frac{\ln c}{\beta}$ just touch line $y=0$ from below. a) $H=0,~\theta=5,~n\rightarrow1$, b) $H=3,~\theta=0.2,~n\rightarrow1$ c) $H=0.25,~\theta=0.2,~n\rightarrow1$ }
\label{figIII}
\end{figure}

\begin{figure}[htb]
\begin{center}
\begin{minipage}{0.32\linewidth}
\includegraphics[width=\linewidth]{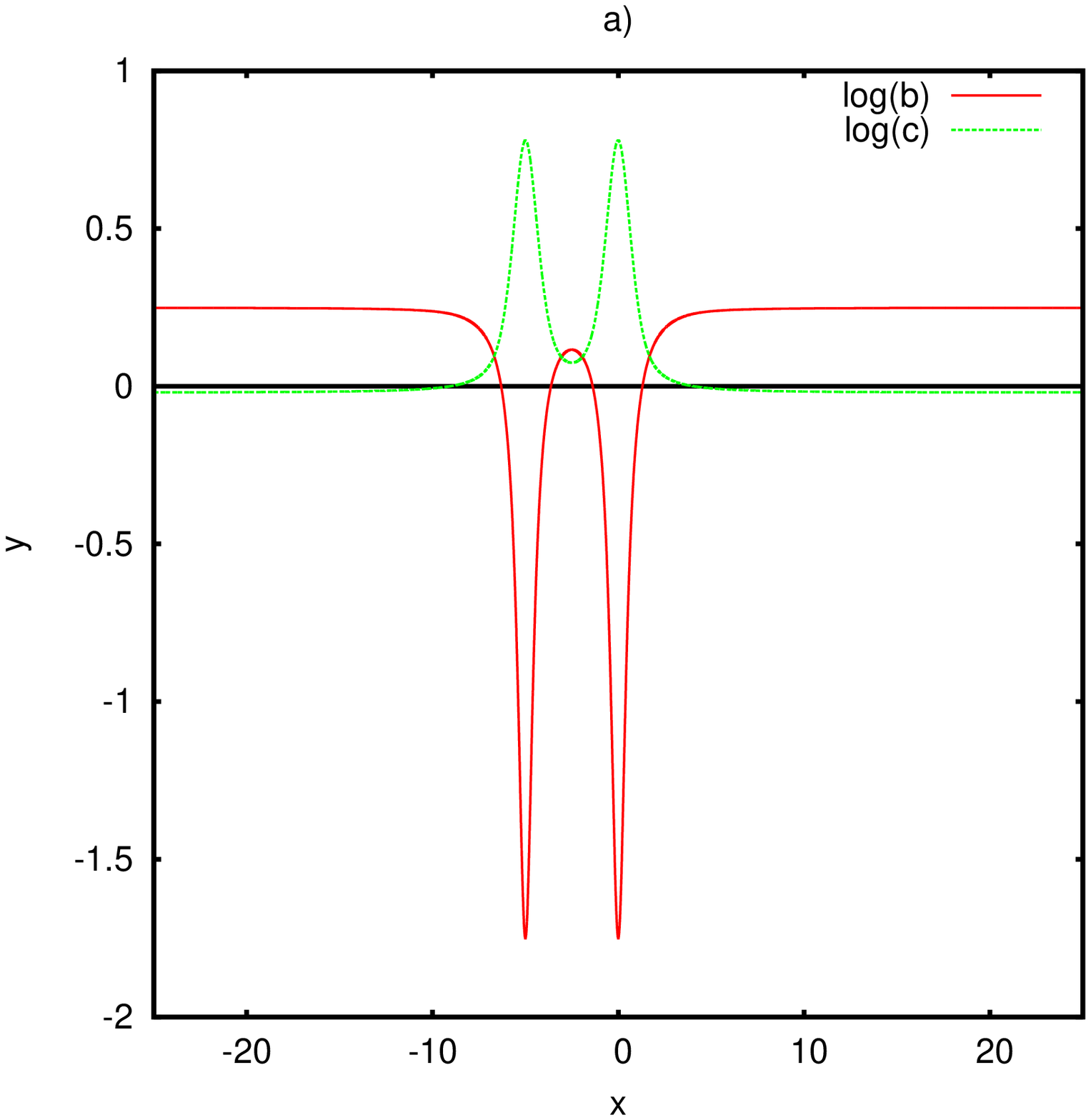}
\end{minipage}%
\begin{minipage}{0.32\linewidth}
\includegraphics[width=\linewidth]{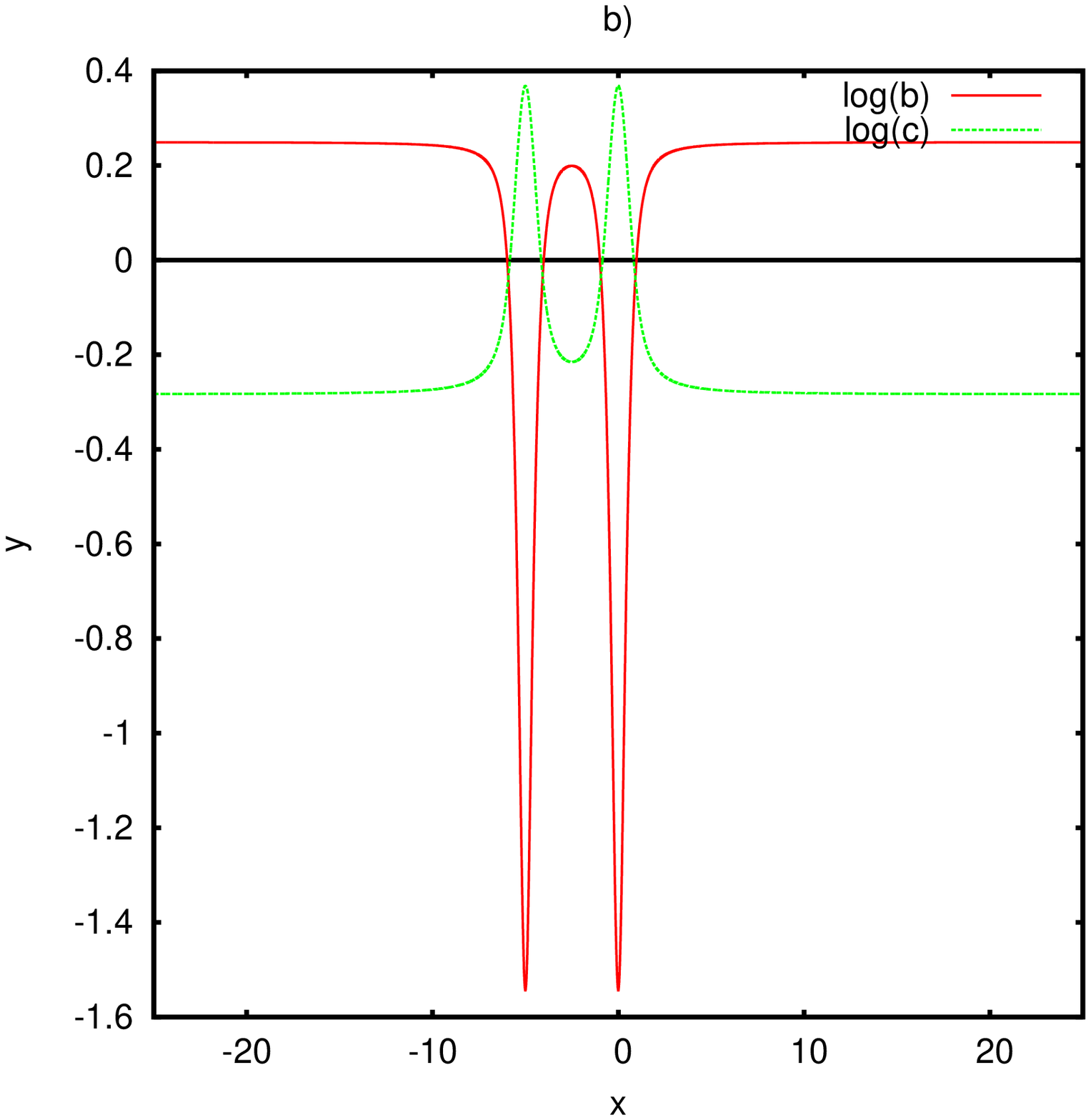}
\end{minipage}%
\begin{minipage}{0.32\linewidth}
\includegraphics[width=\linewidth]{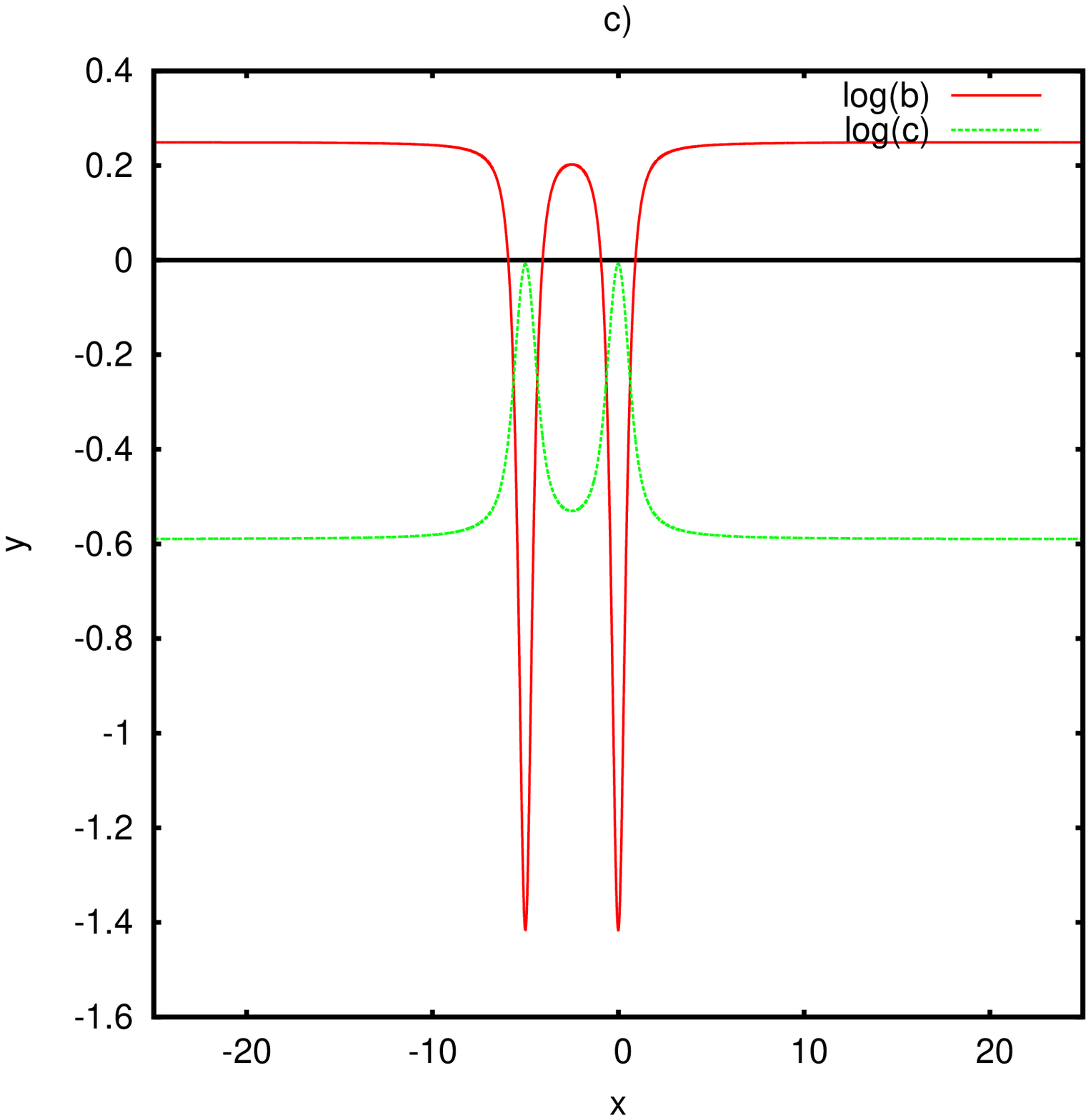}
\end{minipage}
\end{center}
\caption{(Color online) a) Phase IV: Commensurate metallic incommensurate anti-ferromagnetic: $\frac{\ln b}{\beta}$ has two dirac seas and $\frac{\ln c}{\beta}$ has one inverted dirac sea. $H=0.25,~\theta=5,~n=0.25$; b) Phase V: Incommensurate metallic incommensurate anti-ferromagnetic: $\frac{\ln b}{\beta}$ has two dirac seas and $\frac{\ln c}{\beta}$ has two inverted dirac seas. $H=0.25,~\theta=5,~n=0.6$; c) Phase VI: Insulating incommensurate anti-ferromagnetic: $\frac{\ln b}{\beta}$ has two dirac seas and $\frac{\ln c}{\beta}$ just touch the line $y=0$ from below. $H=0.25,~\theta=5,~n\rightarrow1$.}
\label{figV}
\end{figure}

\begin{figure}[htb]
\begin{center}
\begin{minipage}{0.35\linewidth}
\includegraphics[width=\linewidth]{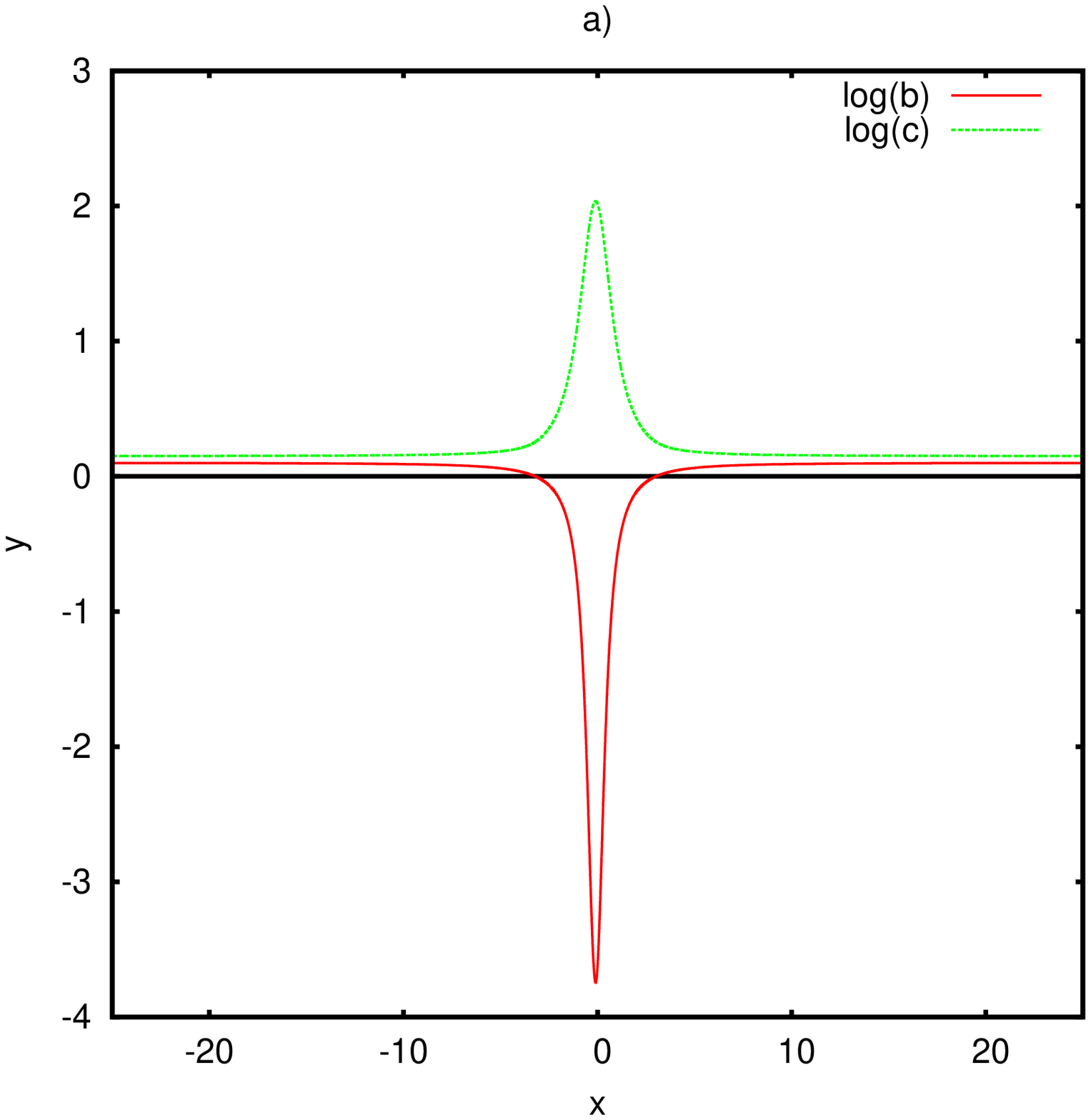}
\includegraphics[width=\linewidth]{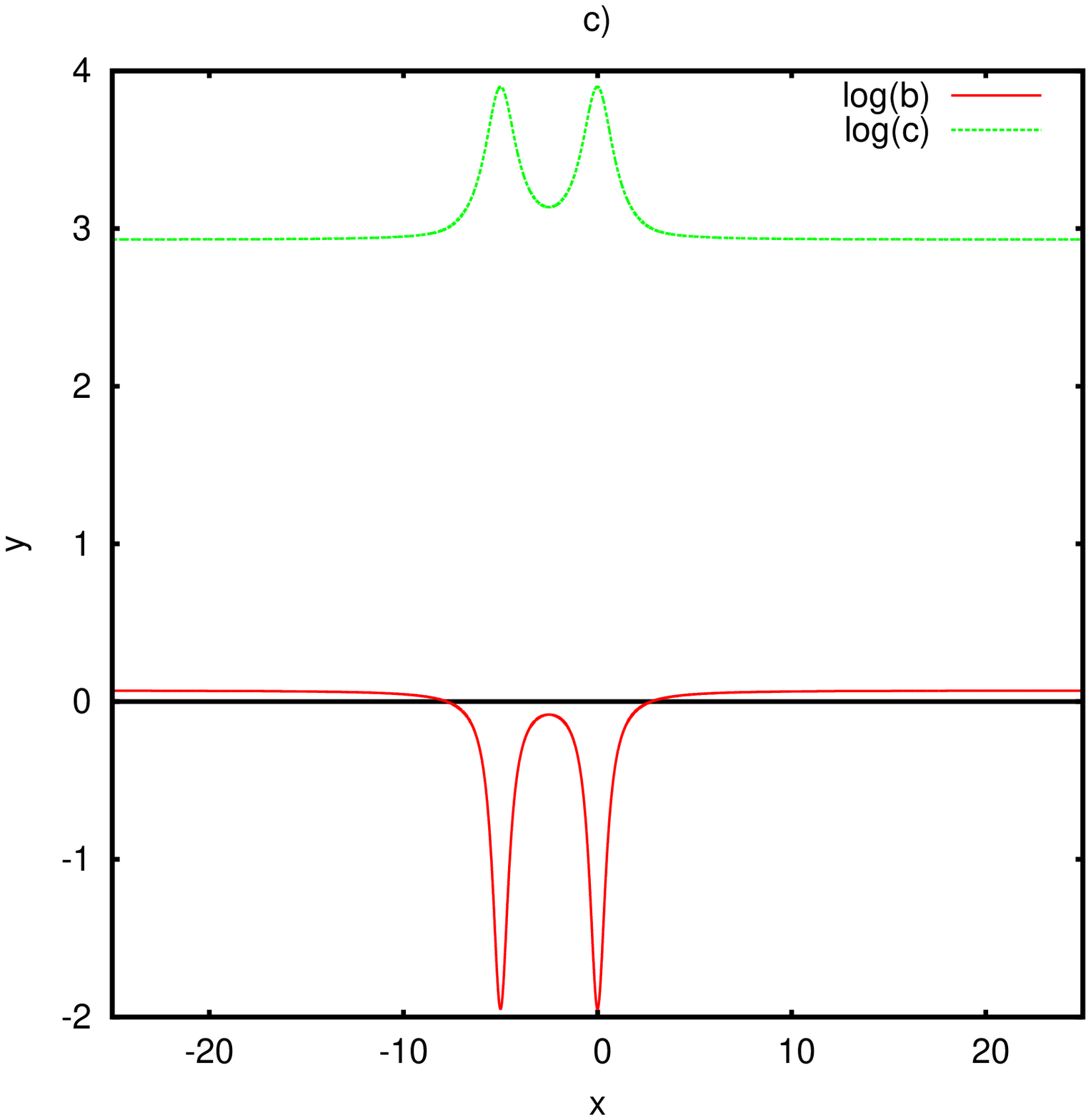}
\end{minipage}
\begin{minipage}{0.35\linewidth}
\includegraphics[width=\linewidth]{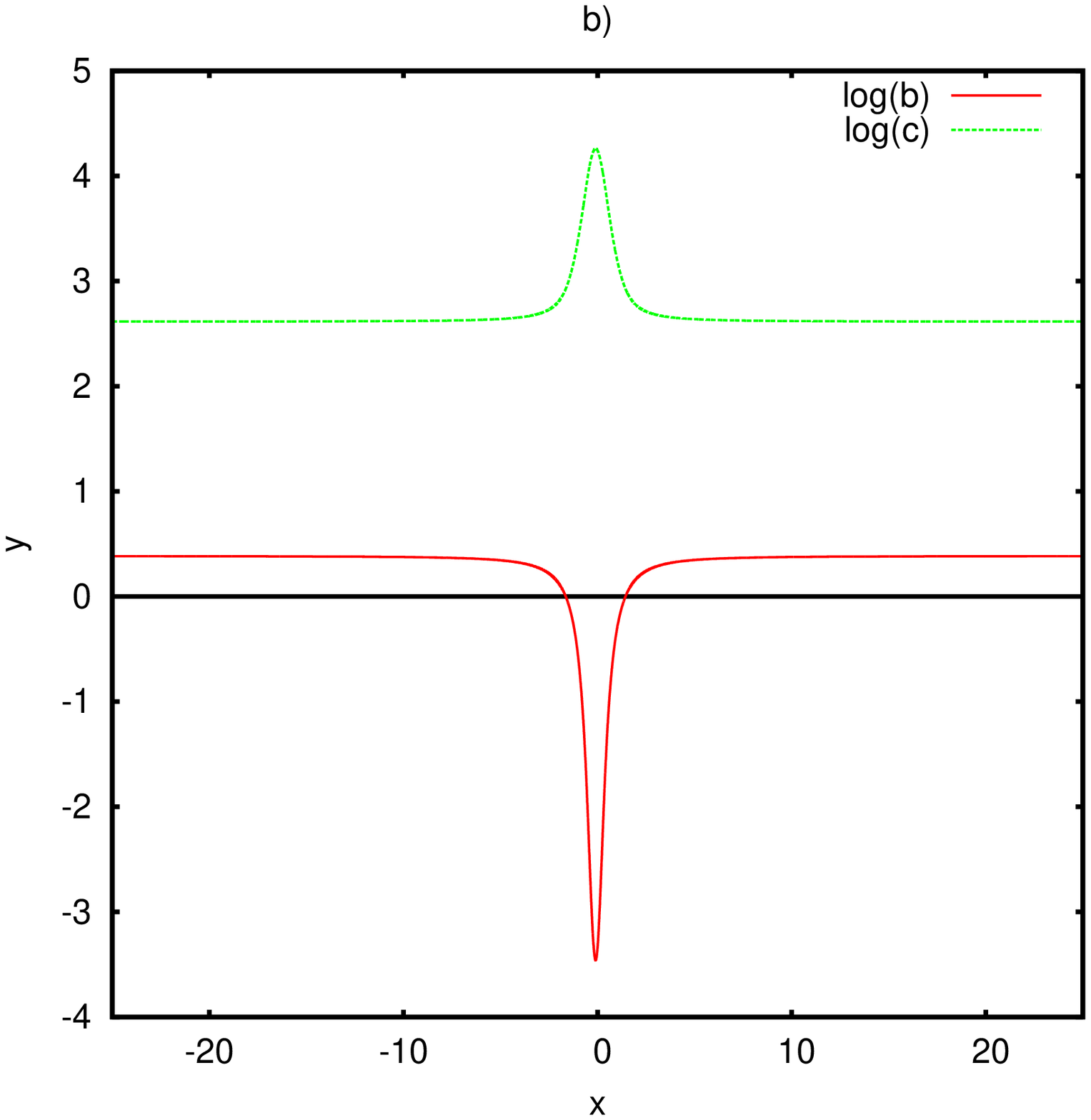}
\includegraphics[width=\linewidth]{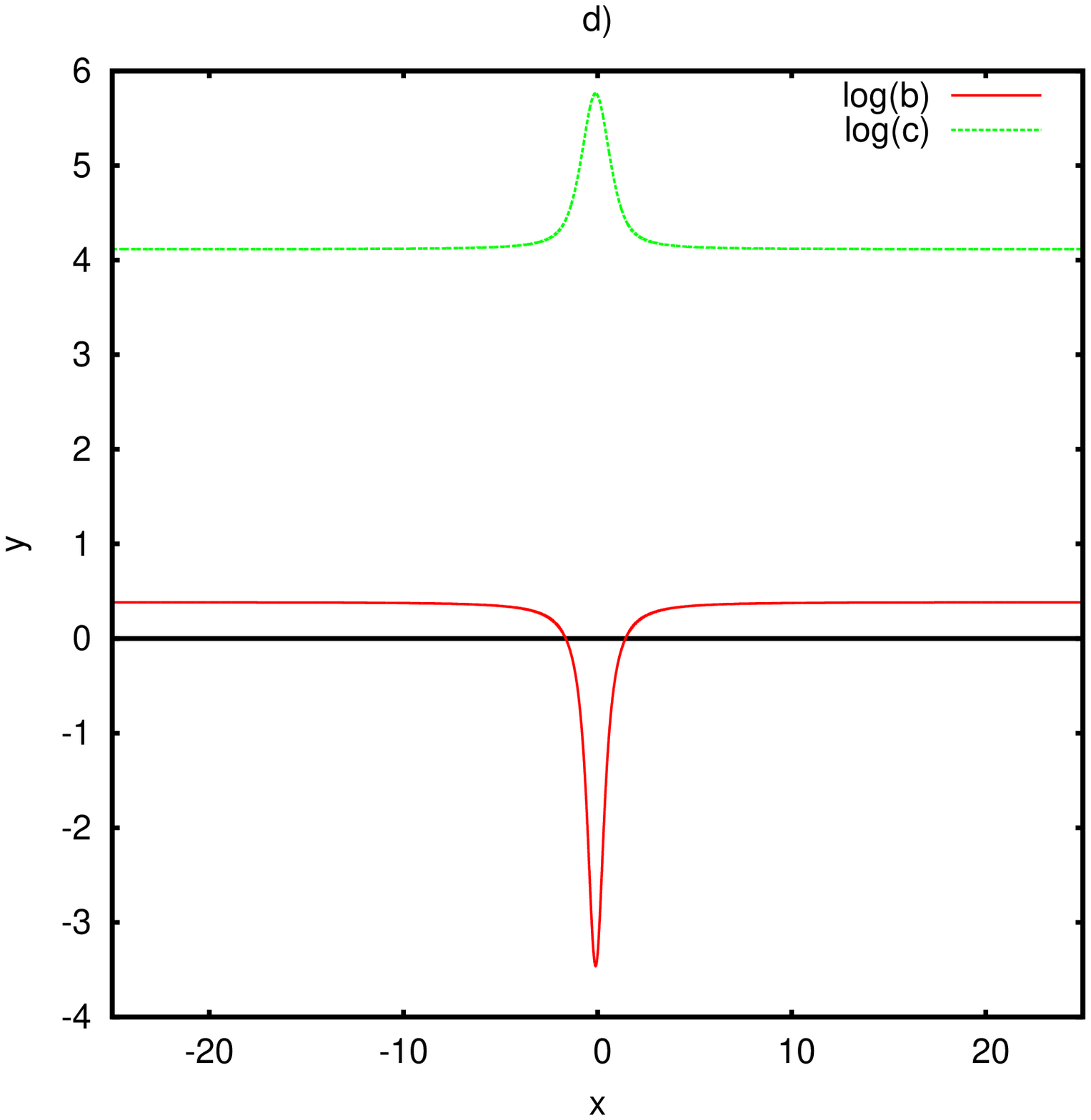}
\end{minipage}
\end{center}
\caption{(Color online) Phase VII: Commensurate metallic ferromagnetic: $\frac{\ln b}{\beta}$ has one dirac sea and $\frac{\ln c}{\beta}$ is completely above the line $y=0$. a) $H=0.25,~\theta=0.2,~n=0.1$, b) $H=3,~\theta=0.2,~n=0.2$, c) $H=3,~\theta=5,~n=0.075$ d) $H=4.5,~\theta=0.2,~n=0.2$. }
\label{figVII}
\end{figure}

\begin{figure}[htb]
\begin{center}
\begin{minipage}{0.32\linewidth}
\includegraphics[width=\linewidth]{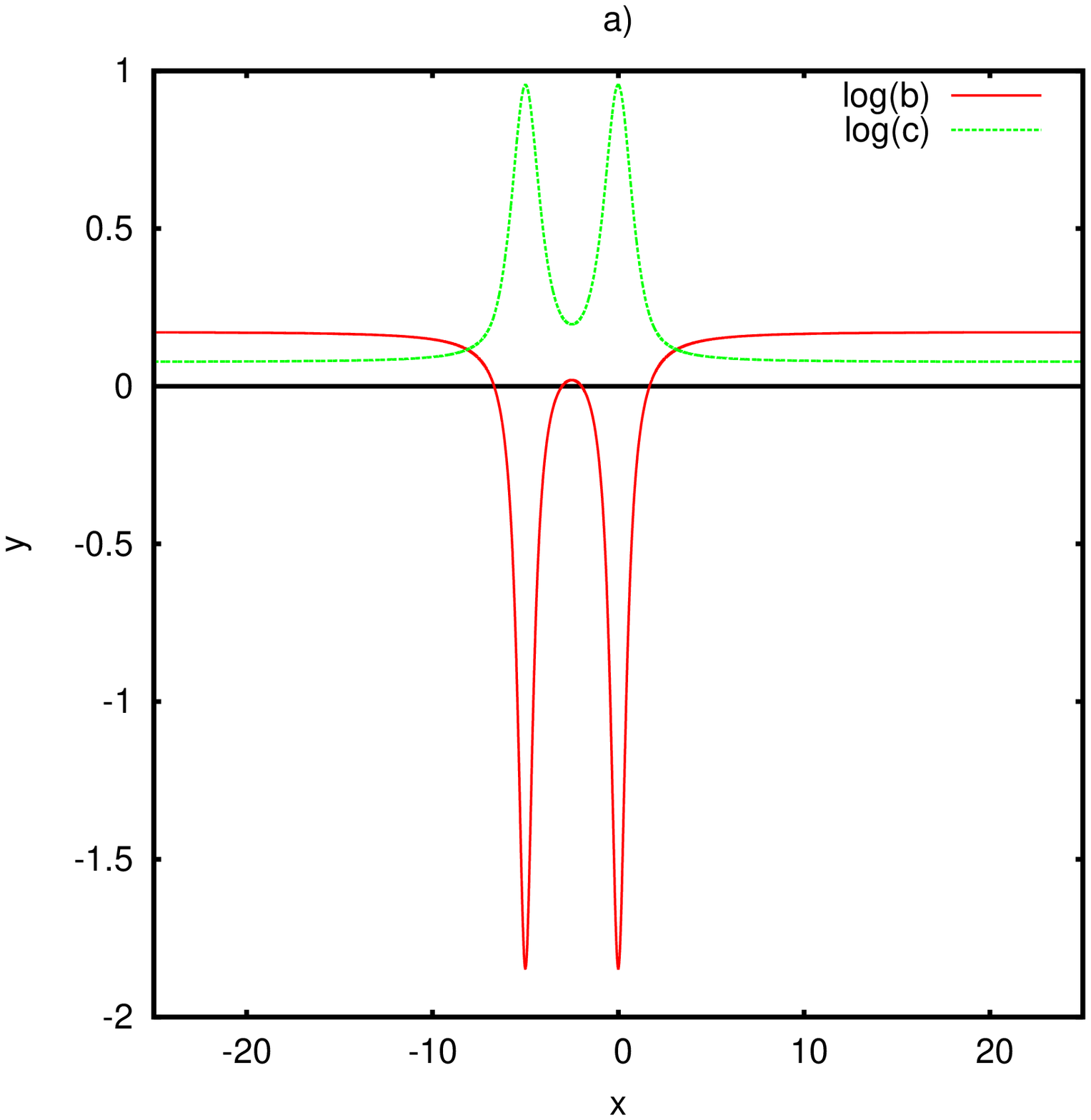}
\end{minipage}%
\begin{minipage}{0.32\linewidth}
\includegraphics[width=\linewidth]{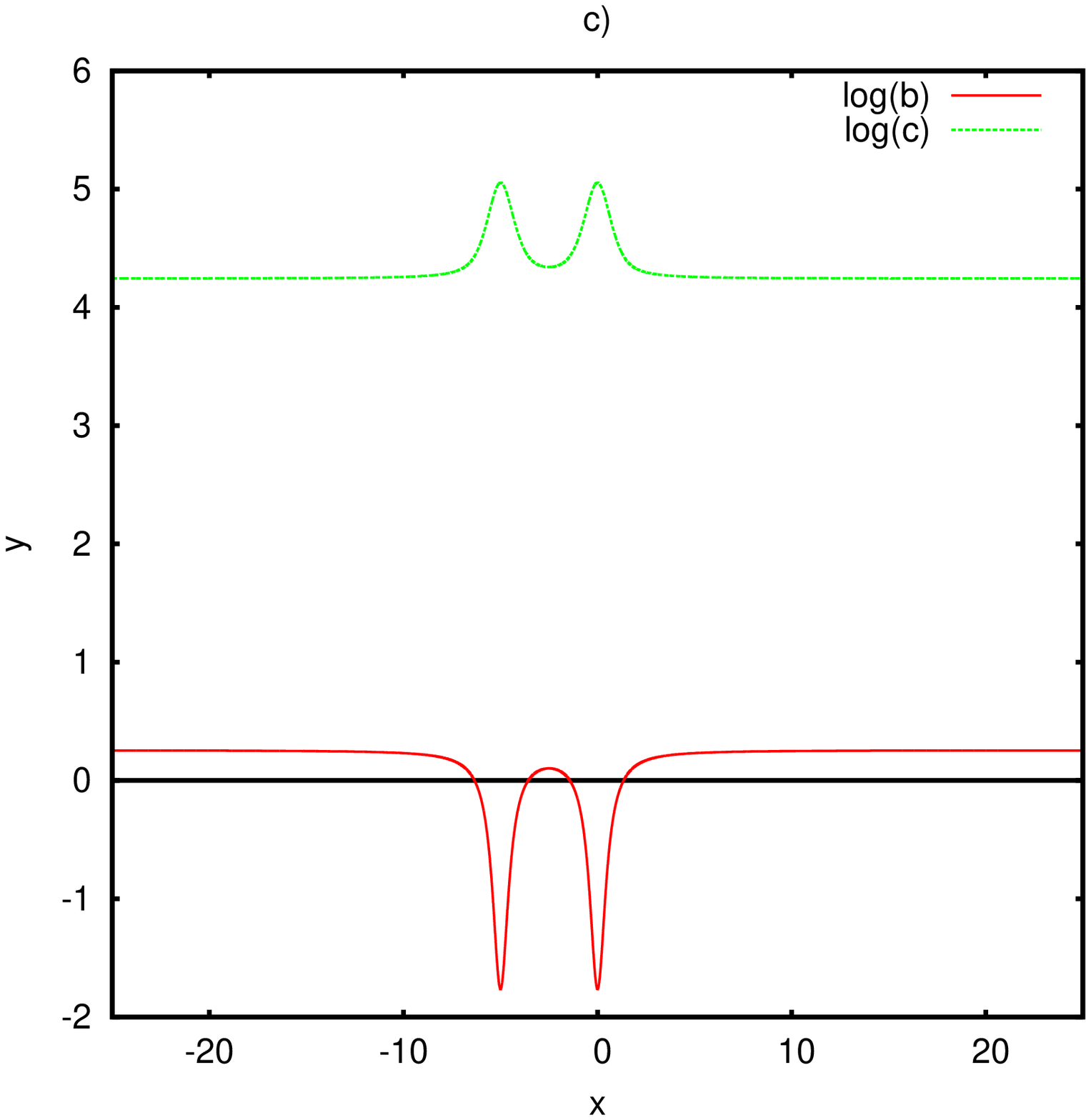}
\end{minipage}%
\begin{minipage}{0.32\linewidth}
\includegraphics[width=\linewidth]{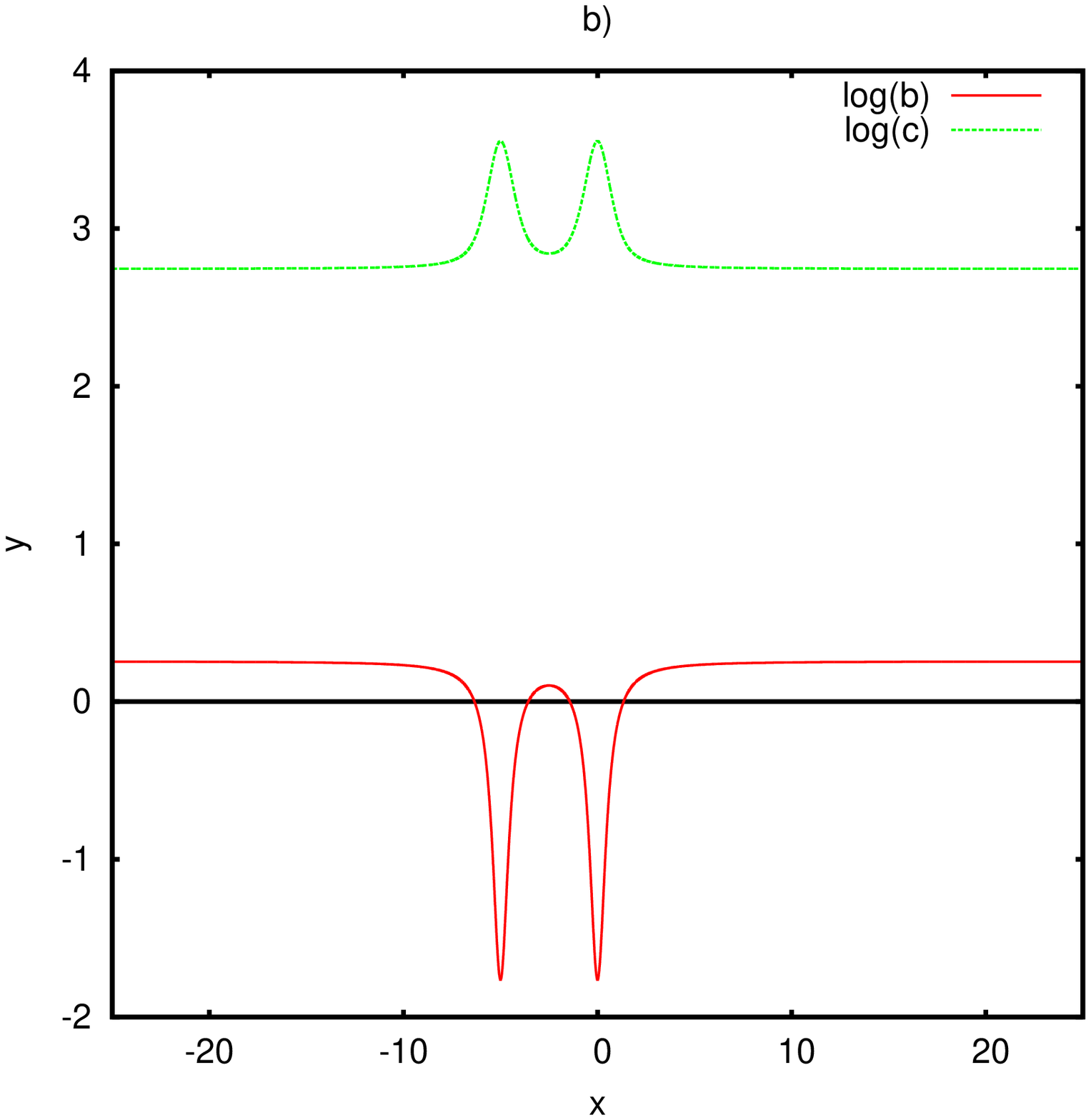}
\end{minipage}
\end{center}
\caption{(Color online) Phase VIII: Incommensurate metallic ferromagnetic: $\frac{\ln b}{\beta}$ has two dirac seas and $\frac{\ln c}{\beta}$ is completely above the line $y=0$. a) $H=0.25,~\theta=5,~n=0.25$, b) $H=3,~\theta=5,~n=0.2$, c) $H=4.5,~\theta=5,~n=0.2$.  }
\label{figVIII}
\end{figure}

\begin{figure}[htb]
\begin{center}
\begin{minipage}{0.35\linewidth}
\includegraphics[width=\linewidth]{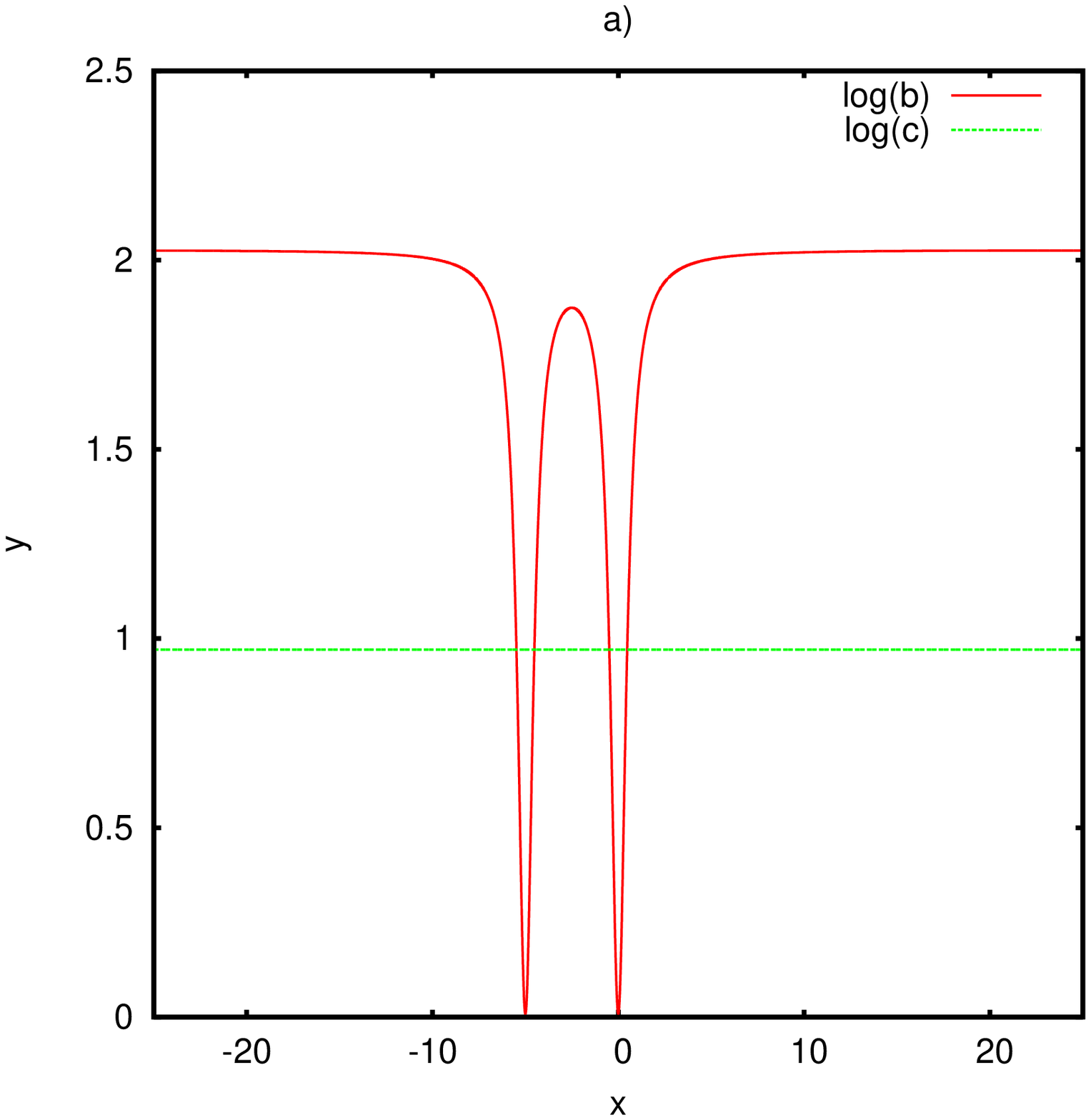}
\end{minipage}
\begin{minipage}{0.35\linewidth}
\includegraphics[width=\linewidth]{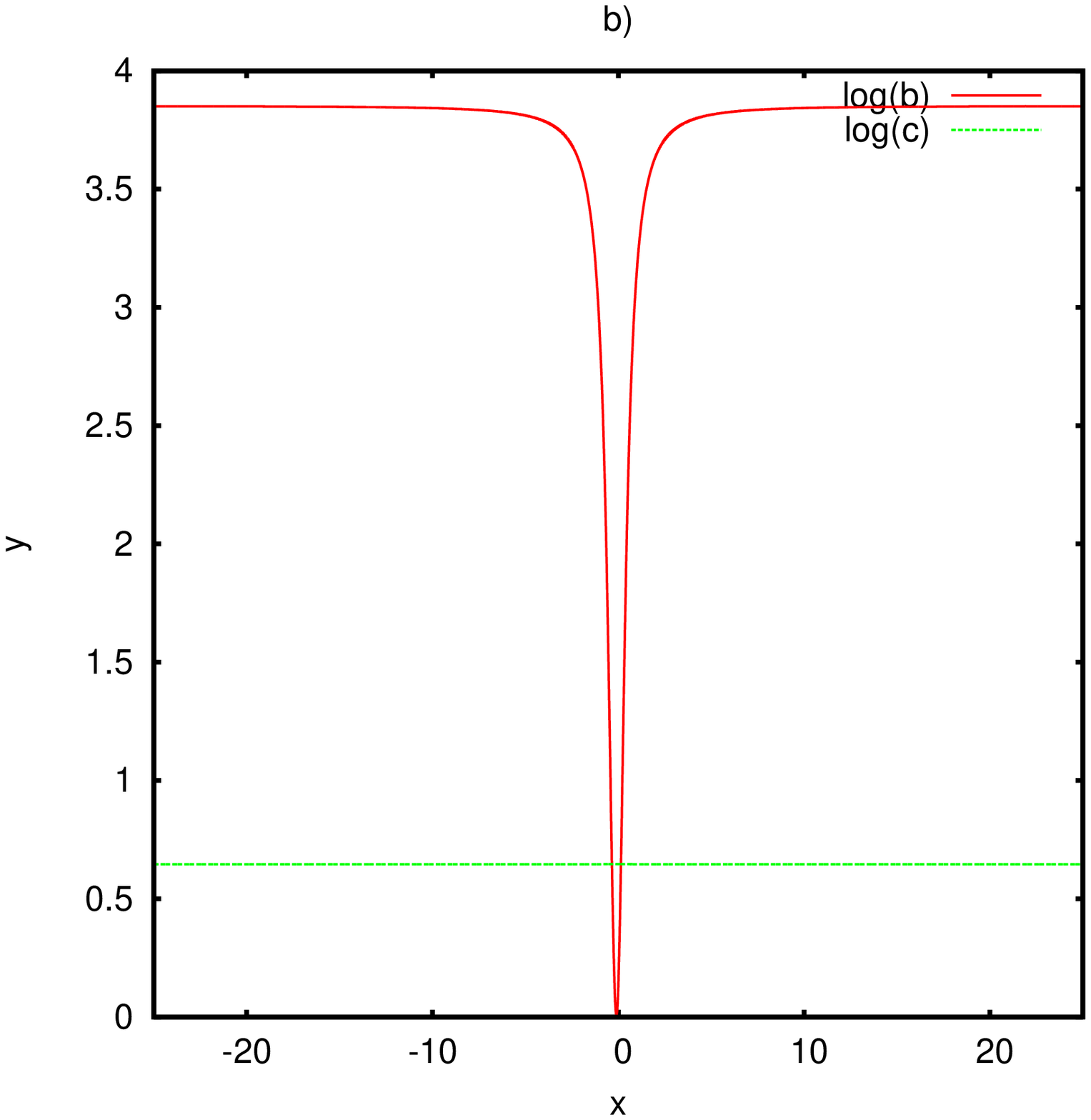}
\end{minipage}
\end{center}
\caption{(Color online) Phase IX: Insulating ferromagnetic: $\frac{\ln b}{\beta}$ just touch the line $y=0$ from above and $\frac{\ln c}{\beta}$ is completely above the line $y=0$. a) $H=3,~\theta=5,~n\rightarrow 1$, b) $H=4.5,~\theta=0.2,~n\rightarrow 1$.  }
\label{figIX}
\end{figure}

\begin{figure}[htb]
\begin{center}
\begin{minipage}{0.35\linewidth}
\includegraphics[width=\linewidth]{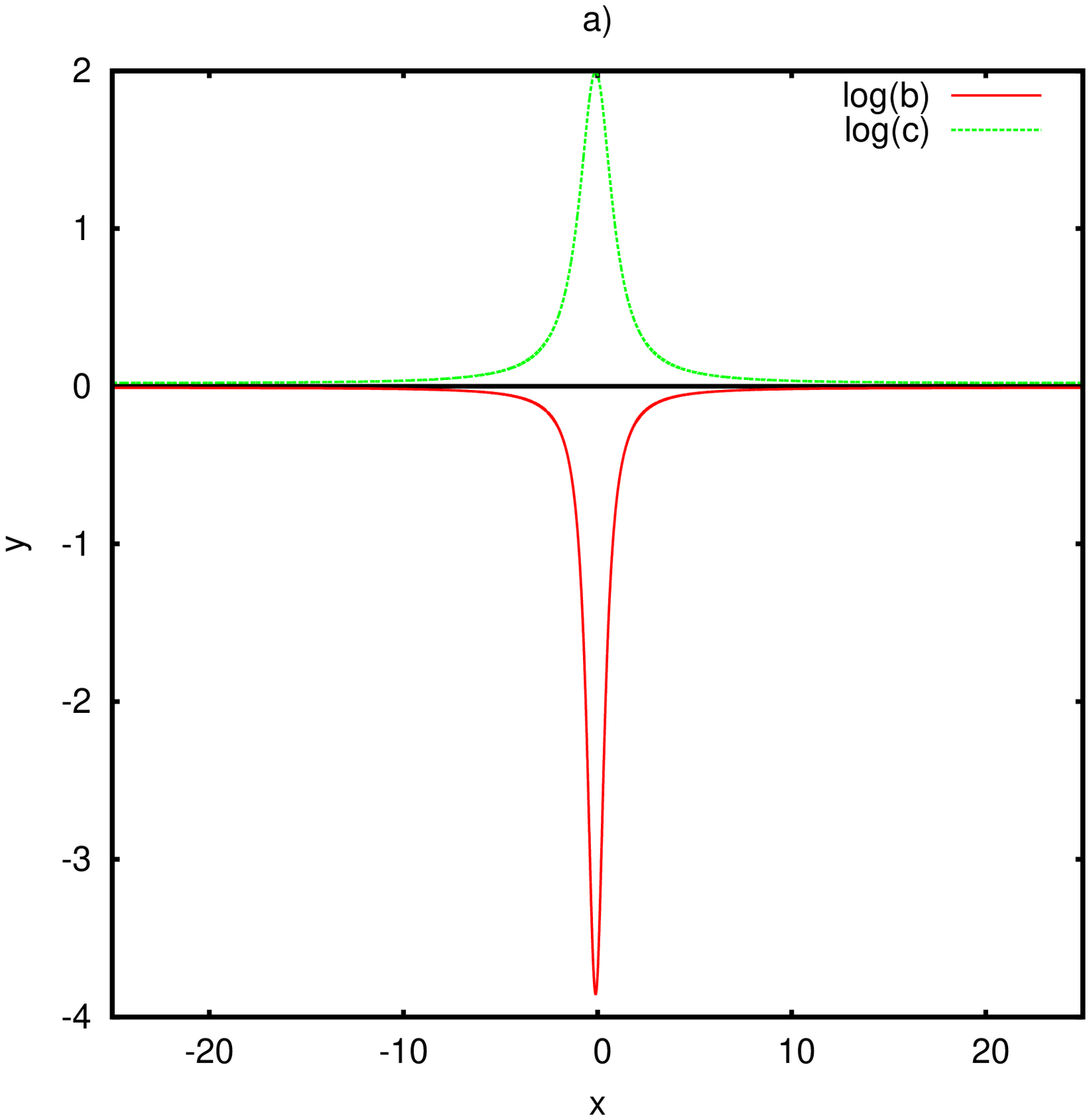}
\includegraphics[width=\linewidth]{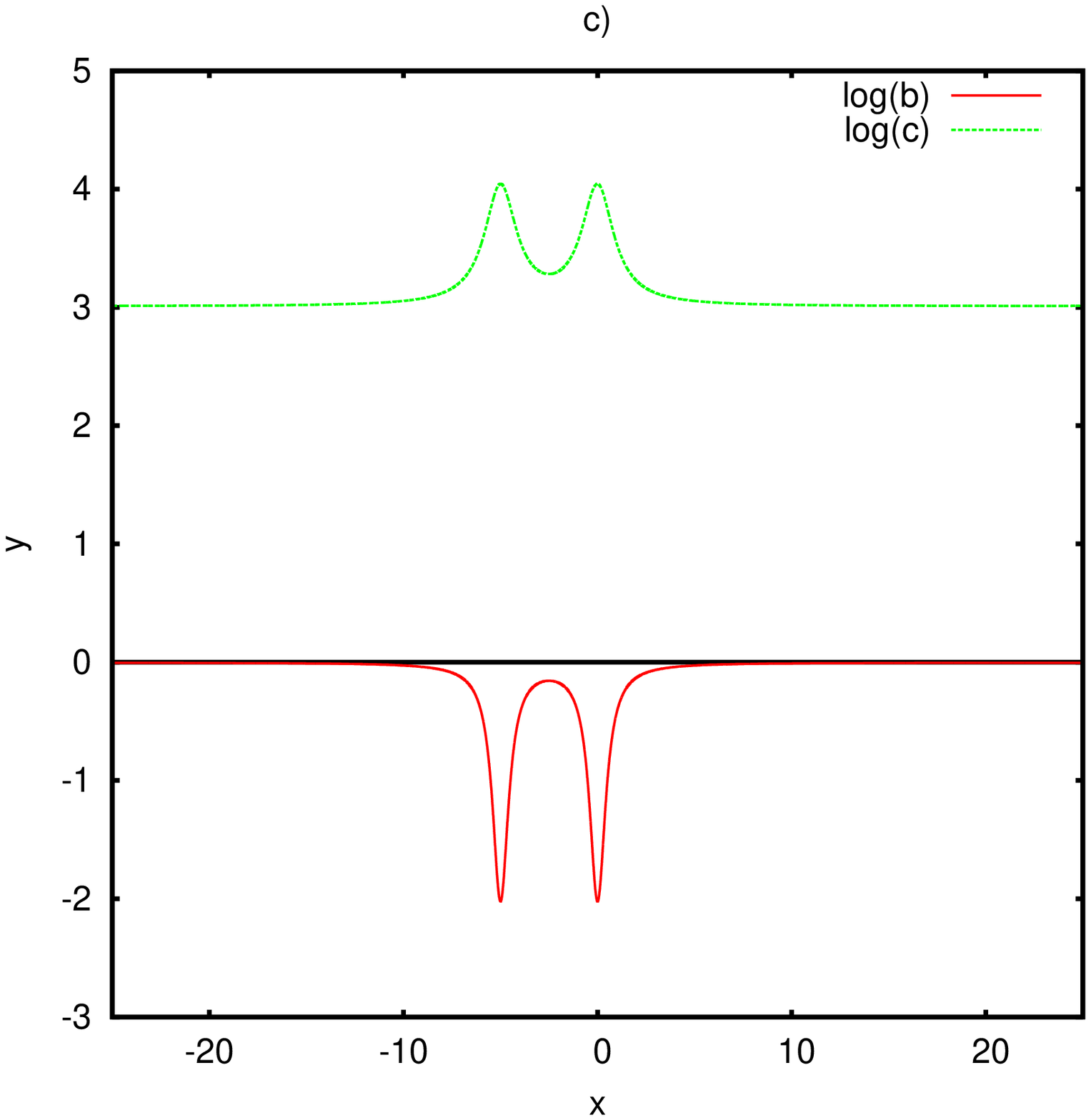}
\end{minipage}
\begin{minipage}{0.35\linewidth}
\includegraphics[width=\linewidth]{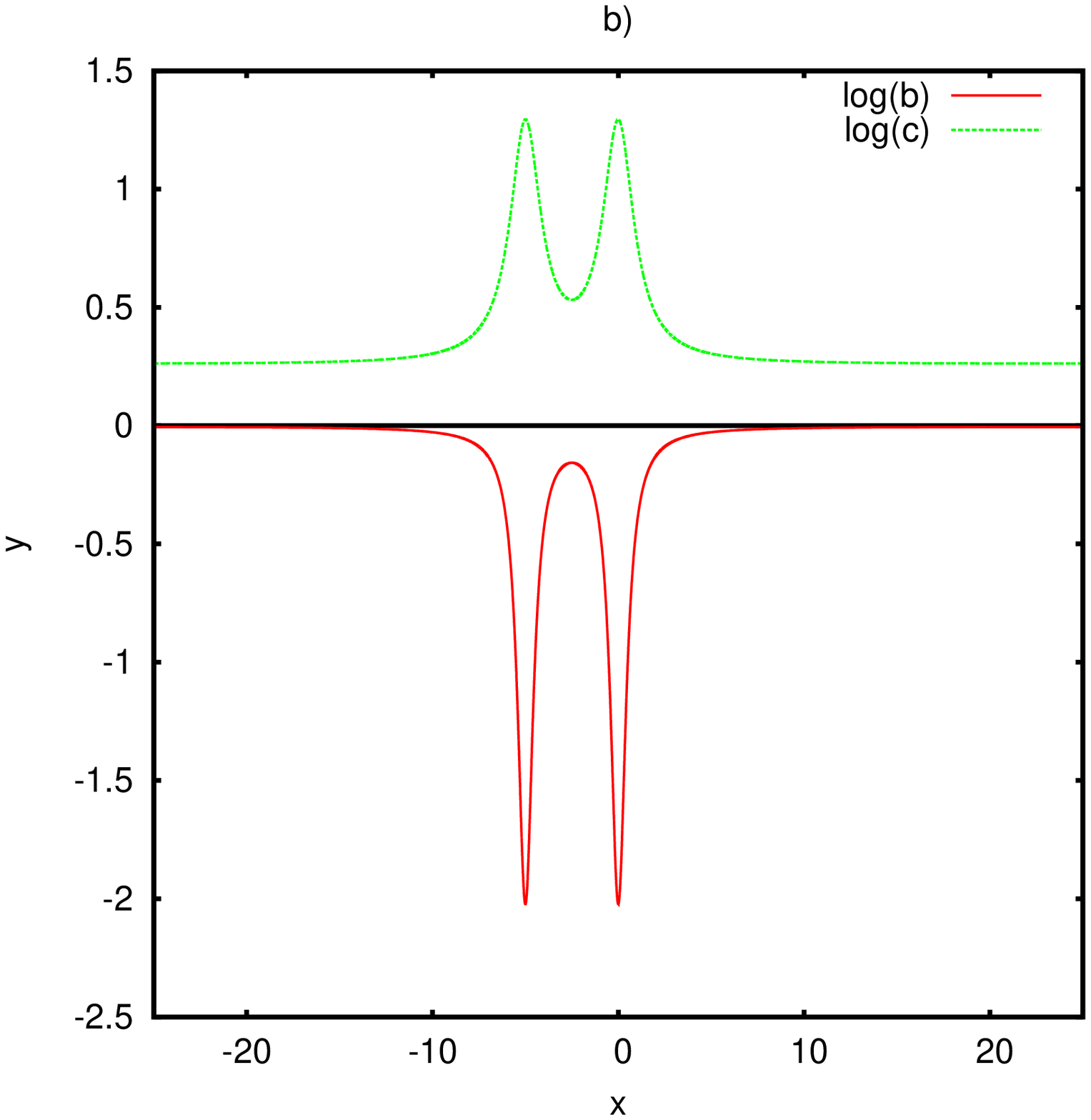}
\includegraphics[width=\linewidth]{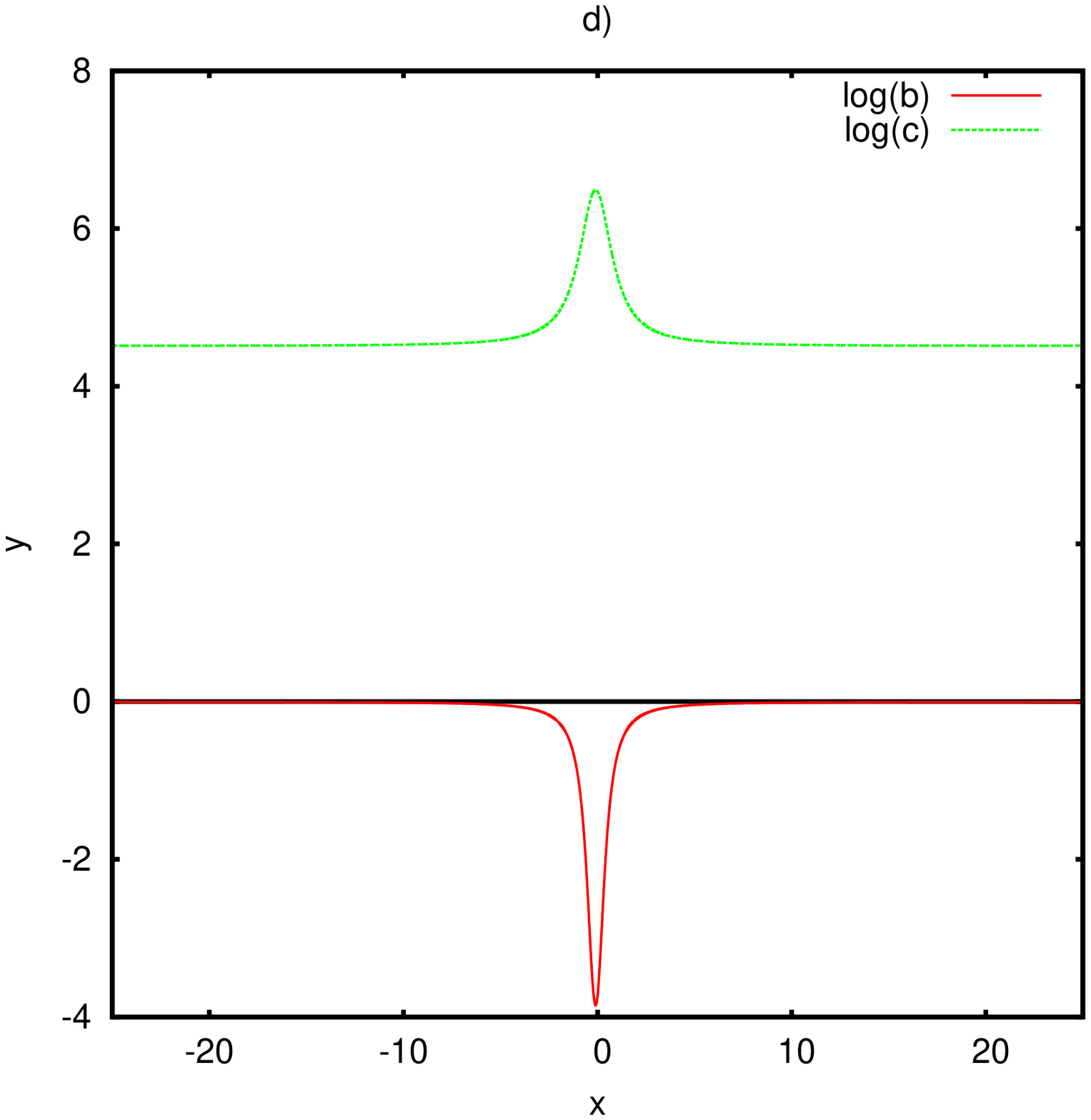}
\end{minipage}
\end{center}
\caption{(Color online) Phase X: zero density phase $n=0$: $\frac{\ln b}{\beta}$ touch the line $y=0$ from below when $x \rightarrow \pm \infty$ while $\frac{\ln c}{\beta}$ is completely above. a) $H=0,~\theta=0.2,~n\rightarrow 0$, b) $H=0.25,~\theta=5,~n\rightarrow 0$, c) $H=3,~\theta=5,~n\rightarrow0$ d) $H=4.5,~\theta=0.2,~n\rightarrow0$. }
\label{figX}
\end{figure}

In Figures \ref{figI} to \ref{figX} we present the auxiliary functions for typical points of each phase in Figures \ref{fig1} and \ref{fig3}. In order to better view the qualitative changes we have highlighted the $y=0$ axis (the vertical axis $y$ represents the auxiliary functions $\frac{\ln b}{\beta}$ and $\frac{\ln c}{\beta}$). This line separates the occupied levels from the not occupied ones. Here one should look at ($\frac{\ln \mathfrak{c}(x)}{\beta}$) $\frac{\ln \mathfrak{b}(x)}{\beta}$ as representing the possible scenarios for the (inverted) Dirac seas. For instance, the transition from phase \textrm{I} to \textrm{II} in phase diagram \ref{fig1} a) is marked by a change from one inverted Dirac sea (Figure \ref{figI}) to two inverted Dirac seas (Figure \ref{figII}) for  $\frac{\ln \mathfrak{c}(x)}{\beta}$, thereby increasing the number of gapless modes. Such discussion has already appeared before in terms of the dressed energy functions in another generalization of the t-J model\cite{TJHIGHER}. 
Therefore, one may identify the transition between a commensurate metallic order to incommensurate metallic order. Likewise, a similar analysis can be performed to every transition line by comparing the behaviour of the auxiliary functions displayed in the Figures \ref{figI} to \ref{figX} as we move from one phase to another.

\end{document}